\documentclass[pdflatex,sn-mathphys-num]{sn-jnl}
\usepackage{graphicx}
\usepackage{multirow}
\usepackage{amsmath,amssymb,amsfonts}
\usepackage{booktabs}
\usepackage{bm}
\usepackage{placeins}
\usepackage{xcolor}
\begin{document}

\title[Merged amplitude encoding for CCQKAN]{Merged amplitude encoding for Chebyshev quantum Kolmogorov--Arnold networks}

\author*[1]{\fnm{Hikaru} \sur{Wakaura}}\email{hikaruwakaura@gmail.com}

\affil*[1]{\orgname{QuantScape Inc.}, \orgaddress{\street{4-11-18, Manshon-Shimizudai}, \city{Meguro}, \postcode{153-0064}, \state{Tokyo}, \country{Japan}}}

\abstract{Quantum Kolmogorov--Arnold networks evaluate each edge activation function as a quantum inner product, creating a trade-off between qubit count and the number of circuit executions per forward pass. We introduce merged amplitude encoding for the Chebyshev-edge classical-to-classical QKAN (CCQKAN), which packs the element-wise products of all $n$ input-edge vectors of an output node into a single amplitude state, reducing circuit executions by a factor of $n$ while using no more qubits than the sequential SWAP-test baseline once its measurement registers are counted. The merged and original circuits compute the same quantity exactly; we characterize what this consolidation costs in trainability. Across 10 network configurations and 16 random seeds, the two circuits are indistinguishable under ideal conditions. Under finite-shot training with a hardware-standard SPSA optimizer at matched measurement budget, the merged circuit incurs a small systematic loss deficit that grows with network width (significant in 6 of 10 configurations), consistent with amplified estimator noise in the merged state; under depolarizing noise no systematic difference is detected. A parameter-matched single-qubit data re-uploading baseline reaches substantially higher loss on the same tasks while using fewer qubits and executions, delineating complementary resource regimes. No quantum advantage is claimed.}


\maketitle

\section{Introduction}
\label{sec:intro}

Quantum machine learning has advanced rapidly as one of the promising near-term applications of quantum computers~\cite{Biamonte2017,Schuld2019,Cerezo2021}.
Variational quantum algorithms~\cite{Cerezo2021} and parameterized quantum circuits~\cite{Benedetti2019} have emerged as leading candidates for the noisy intermediate-scale quantum (NISQ) era~\cite{Preskill2018}, enabling hybrid quantum-classical approaches to classification, regression, and generative modeling.
However, quantum neural network architectures face a trade-off between the number of qubits required and the number of circuit executions per forward pass, and the search for favorable balances between these resources is an active line of work~\cite{PerezSalinas2020,Benedetti2019,Cerezo2021}.

Kolmogorov--Arnold networks (KAN) are function approximators based on the Kolmogorov--Arnold representation theorem~\cite{Kolmogorov1957,Arnold1957}, which states that any multivariate continuous function can be decomposed into compositions and sums of univariate functions.
Recently, Liu et al.~\cite{Liu2024KAN} introduced KAN as a neural network architecture in which learnable activation functions are placed on edges rather than nodes, demonstrating competitive or superior performance compared to multilayer perceptrons in terms of both accuracy and interpretability.

The quantum extension of KAN has been explored in several recent works~\cite{KunduQKAN2024,WakauraVQKAN2026,AdaptiveVQKAN2025,EnhancedVQKAN2025,QuKAN2025}.
KAN architectures have also been employed \emph{within} quantum workflows---for example, as the function approximator in reinforcement-learning-based quantum architecture search~\cite{KANQAS2024}---which is distinct from realizing the KAN itself on a quantum circuit.
In particular, the classical-to-classical QKAN framework (CCQKAN)~\cite{GonzalezOtero2025} realizes Chebyshev-basis edge functions of degree $d$ with learnable coefficients, implemented there through quantum-singular-value-transformation block encodings with a Hadamard-test readout.
In this work we adopt the same Chebyshev edge functions but a simpler statevector-level realization: the basis vector $\bm{T}(x)$ and the coefficient vector $\bm{c}$ are encoded as quantum amplitude states and their inner product is evaluated via a SWAP test~\cite{Buhrman2001}.
The CCQKAN architecture admits two extreme execution strategies for a two-layer $[n,n,1]$ network with Chebyshev degree $d$: (a) \emph{fully parallel} evaluation, which allocates a separate qubit register to each of the $(n^2+n)$ edges and requires $Q_{\mathrm{par}} = (n^2+n)\lceil\log_2(d+1)\rceil$ qubits but only one circuit execution per layer; and (b) \emph{fully sequential} evaluation, which reuses a single $Q_{\mathrm{seq}} = \lceil\log_2(d+1)\rceil$-qubit register but requires $(n^2+n)$ separate circuit executions.
For instance, a $[4,4,1]$ network with $d=5$ requires either 60 qubits (parallel) or 3 qubits with 20 separate circuit calls per layer (sequential).

Between these extremes, alternative encoding strategies that batch multiple edge evaluations into a single circuit may offer a more favorable trade-off.
Data re-uploading circuits~\cite{PerezSalinas2020} have demonstrated that encoding more classical information into fewer qubits (at the cost of circuit depth) can be a favorable trade-off; the present work pursues a related but distinct strategy---consolidating information from multiple edges into a single amplitude state.

In this paper, we present merged amplitude encoding for CCQKAN.
Rather than evaluating each edge independently, this technique packs the element-wise products of Chebyshev coefficient and basis vectors for all $n$ input edges of a given output node into a single amplitude state, computing their sum in one circuit execution using $Q_{\mathrm{red}} = \lceil\log_2(n(d+1))\rceil$ qubits.
Compared to the sequential baseline (which is the practically relevant alternative), the merged approach reduces circuit executions by a factor of $n$ while using no more qubits once the baseline's SWAP-test registers are counted ($\Delta Q = Q_{\mathrm{red}} - Q_{\mathrm{seq}}$ is $1$--$2$ without measurement overhead and $\leq 0$ with it; Table~\ref{tab:qubits}), though each merged execution uses $O(n)$ times more gates, leaving the total gate count approximately unchanged (Sec.~\ref{sec:resource}).
Since the merged encoding computes the same mathematical quantity as the original---the sum of edge activations [Eq.~\eqref{eq:sum_identity}]---the two architectures are functionally equivalent, differing only in how the computation is distributed across circuit executions.
This mathematical equivalence does not, however, guarantee that the two circuits have identical trainability or noise sensitivity when used within an optimization loop.
The main contribution of this work is to characterize empirically, through systematic numerical experiments, when the merged encoding preserves trainability: exactly under ideal conditions, and with a small, $n$-dependent deficit under finite-shot estimation at fixed per-execution shot count.
Here, ``trainability'' refers specifically to the ability of the merged circuit to reach comparable final loss values as the original circuit within the same short optimization budget (20 Adam steps; see Sec.~\ref{sec:experiments}), rather than a formal guarantee on gradient norms, convergence rates, or asymptotic performance.
We compare three circuit variants---original, reduced with parameter transfer (Red-T), and reduced with independent initialization (Red-I)---under ideal, shot-noise, and shot-plus-device-noise conditions, employing Wilcoxon signed-rank tests with 16 seeds for statistical rigor.
While the mathematical identity underlying the merged encoding is elementary, the trainability question is non-trivial: different circuit structures can exhibit different gradient landscapes under noise, and the interaction between amplitude encoding depth and optimization dynamics is not predictable a priori.
We additionally report MNIST digit classification results---both binary (0 vs 1) and 10-class (one-vs-all)---as supplementary consistency checks on real data.

\section{Background}
\label{sec:background}

\subsection{Kolmogorov--Arnold representation}

The Kolmogorov--Arnold representation theorem~\cite{Kolmogorov1957,Arnold1957} states that any continuous function $f: [0,1]^n \to \mathbb{R}$ can be written as
\begin{equation}
  f(\bm{x}) = \sum_{q=0}^{2n} \Phi_q\!\left(\sum_{p=1}^{n} \phi_{q,p}(x_p)\right),
  \label{eq:ka_theorem}
\end{equation}
where $\phi_{q,p}: [0,1]\to\mathbb{R}$ and $\Phi_q:\mathbb{R}\to\mathbb{R}$ are continuous univariate functions.
KAN~\cite{Liu2024KAN} implements a practical variant of this decomposition as a layered network in which each edge carries a learnable univariate function.

\subsection{Chebyshev-edge QKAN (CCQKAN)}
\label{sec:ccqkan}

Following the CCQKAN framework~\cite{GonzalezOtero2025}, each edge activation function $\varphi(x)$ is expanded in the Chebyshev basis of degree $d$:
\begin{equation}
  \varphi(x) = \sum_{k=0}^{d} c_k\, T_k(x),
  \label{eq:cheb_expansion}
\end{equation}
where $T_k(x)$ denotes the Chebyshev polynomial of the first kind of degree $k$ and $\{c_k\}_{k=0}^{d}$ are learnable coefficients.
Evaluating Eq.~\eqref{eq:cheb_expansion} is equivalent to computing the inner product $\langle \bm{c}, \bm{T}(x)\rangle$, where $\bm{c} = (c_0, \ldots, c_d)^\top$ and $\bm{T}(x) = (T_0(x), \ldots, T_d(x))^\top$.

In the quantum implementation, both $\bm{c}$ and $\bm{T}(x)$ are padded to the nearest power of two and normalized, yielding amplitude states $|\hat{\bm{c}}\rangle$ and $|\hat{\bm{T}}\rangle$ in $q_e = \lceil\log_2(d+1)\rceil$-qubit registers.
The inner product is estimated via a SWAP test~\cite{Buhrman2001}: an ancilla qubit is prepared in $|+\rangle$, a controlled-SWAP is applied between the two registers, and the ancilla is measured.
The probability of measuring $|0\rangle$ on the ancilla is
\begin{equation}
  P(0) = \frac{1 + |\langle \hat{\bm{c}} | \hat{\bm{T}} \rangle|^2}{2},
  \label{eq:swap_test}
\end{equation}
from which $|\langle \hat{\bm{c}} | \hat{\bm{T}} \rangle|^2$ is estimated as $2P(0) - 1$.
The un-normalized inner product is then recovered by multiplying by $\|\bm{c}\| \cdot \|\bm{T}(x)\|$.
When using exact statevector simulation, the overlap $\langle \hat{\bm{c}} | \hat{\bm{T}} \rangle$ is computed directly without the SWAP test ancilla.
(The merged encoding introduced in Sec.~\ref{sec:method} replaces the SWAP test with a simpler overlap measurement against the uniform state.)
This amplitude-encoding realization is our own simplification of the Chebyshev edge: the reference implementation of Ref.~\cite{GonzalezOtero2025} evaluates the same edge polynomials through QSVT block encodings, a linear combination of unitaries over polynomial degrees, and a Hadamard-test readout.
The two realizations compute the same edge function; all resource accounting in this paper refers to the amplitude-encoding realization.

For a two-layer $[n,n,1]$ KAN, the first layer contains $n^2$ edges and the second layer $n$ edges, with an affine output transformation (the trainable output-domain rescaling of Ref.~\cite{GonzalezOtero2025}) consisting of learnable scale $\alpha$ and shift $\beta$: $y = \alpha \cdot y_{\mathrm{raw}} + \beta$.
In parallel mode, the total qubit count is
\begin{equation}
  Q_{\mathrm{par}} = (n^2+n)\,q_e = (n^2+n)\,\lceil\log_2(d+1)\rceil.
  \label{eq:qubits_orig}
\end{equation}

The network forward pass for a $[n,n,1]$ architecture proceeds as follows:
\begin{enumerate}
  \item Inputs $\bm{x} \in \mathbb{R}^n$ are normalized from the data range to $[-1,1]$.
  \item For each hidden node $j \in \{1,\ldots,n\}$, the hidden activation is
  \begin{equation}
    h_j = \tanh\!\left(\sum_{i=1}^{n} \varphi_{ij}(x_i)\right),
    \label{eq:hidden_activation}
  \end{equation}
  where $\varphi_{ij}(x_i) = \sum_{k=0}^{d} c_{ij,k} T_k(x_i)$ is evaluated via the quantum inner product.
  In the original circuit, this requires $n$ separate inner product evaluations per hidden node.
  \item The output is $y = \alpha\left(\sum_{i=1}^{n} \varphi_{i}^{(2)}(h_i)\right) + \beta$, computed by $n$ inner products followed by the classical affine output transformation.
\end{enumerate}

\section{Merged amplitude encoding}
\label{sec:method}

\subsection{Derivation}
\label{sec:derivation}

Consider the computation of hidden node $j$ in the first layer.
In the original architecture, each edge $(i,j)$ computes the inner product $\langle \hat{\bm{c}}_{ij}, \hat{\bm{T}}(x_i)\rangle$ between normalized amplitude states, then rescales by $\|\bm{c}_{ij}\|\cdot\|\bm{T}(x_i)\|$ to recover $\varphi_{ij}(x_i) = \sum_k c_{ij,k} T_k(x_i)$.
The hidden node activation is
\begin{equation}
  h_j = \tanh\!\left(\sum_{i=1}^{n} \varphi_{ij}(x_i)\right) = \tanh\!\left(\sum_{i=1}^{n} \sum_{k=0}^{d} c_{ij,k}\, T_k(x_i)\right).
  \label{eq:hidden_node}
\end{equation}
Rather than computing each edge's inner product on a separate qubit register, we define a merged vector
\begin{equation}
  \bm{m}_j = \bigoplus_{i=1}^{n} \left(\bm{c}_{ij} \odot \bm{T}(x_i)\right) \in \mathbb{R}^{n(d+1)},
  \label{eq:merged_vector}
\end{equation}
where $\odot$ denotes element-wise (Hadamard) multiplication and $\oplus$ denotes vector concatenation (not direct sum).
The $l$-th component of $\bm{m}_j$ is explicitly
\begin{equation}
  [\bm{m}_j]_{(i-1)(d+1)+k} = c_{ij,k}\, T_k(x_i),
  \label{eq:merged_components}
\end{equation}
for $i=1,\ldots,n$ and $k=0,\ldots,d$.
The sum of all components therefore equals the desired pre-activation:
\begin{equation}
  S_j \equiv \sum_{l=0}^{n(d+1)-1} [\bm{m}_j]_l = \sum_{i=1}^{n}\sum_{k=0}^{d} c_{ij,k}\, T_k(x_i).
  \label{eq:sum_identity}
\end{equation}
To evaluate $S_j$ quantumly, we pad $\bm{m}_j$ with zeros to dimension $D = 2^{\lceil\log_2(n(d+1))\rceil}$ and prepare the normalized amplitude state
\begin{equation}
  |\tilde{\bm{m}}_j\rangle = \frac{\bm{m}_j^{\mathrm{pad}}}{\|\bm{m}_j^{\mathrm{pad}}\|} = \frac{\bm{m}_j^{\mathrm{pad}}}{\|\bm{m}_j\|},
  \label{eq:merged_state}
\end{equation}
where the last equality holds because the zero-padded entries do not contribute to the norm.
The key observation is that the sum of all amplitudes of a quantum state equals $\sqrt{D}$ times its overlap with the uniform superposition.
Exploiting this, we compute the overlap with the uniform state $|U\rangle = D^{-1/2}\sum_{l=0}^{D-1}|l\rangle$:
\begin{align}
  \langle U | \tilde{\bm{m}}_j \rangle
  &= \frac{1}{\sqrt{D}} \sum_{l=0}^{D-1} \frac{[\bm{m}_j^{\mathrm{pad}}]_l}{\|\bm{m}_j\|} \nonumber\\
  &= \frac{1}{\sqrt{D}\,\|\bm{m}_j\|} \sum_{l=0}^{n(d+1)-1} [\bm{m}_j]_l \nonumber\\
  &= \frac{S_j}{\sqrt{D}\,\|\bm{m}_j\|},
  \label{eq:overlap_derivation}
\end{align}
and thus
\begin{equation}
  S_j = \|\bm{m}_j\| \sqrt{D}\, \langle U | \tilde{\bm{m}}_j \rangle.
  \label{eq:Sj_recovery}
\end{equation}
This identity is exact; the merged encoding introduces no approximation beyond that inherent in the amplitude encoding and inner product estimation.
We emphasize that the construction above is mathematically elementary---it follows directly from the linearity of the inner product and the definition of amplitude encoding.
The contribution of this work is therefore not the identity itself, but rather two-fold: (1)~identifying that the specific structure of the CCQKAN forward pass permits this batching (which is not obvious for arbitrary quantum neural network architectures), and (2)~the systematic empirical verification that rearranging the computation preserves trainability when the circuit is used within a gradient-based optimization loop (Sec.~\ref{sec:results}).

Since $|U\rangle = H^{\otimes Q_{\mathrm{red}}}|0\rangle^{\otimes Q_{\mathrm{red}}}$ (the uniform state is prepared by applying Hadamard gates to the all-zero state), the overlap $\langle U | \tilde{\bm{m}}_j \rangle$ can be evaluated without a SWAP test: one applies $H^{\otimes Q_{\mathrm{red}}}$ to $|\tilde{\bm{m}}_j\rangle$ and measures the probability of the all-zero outcome $|0\rangle^{\otimes Q_{\mathrm{red}}}$, which equals $|\langle U | \tilde{\bm{m}}_j \rangle|^2$.
In ideal simulation, the overlap is computed directly from the statevector.
Under shot-based simulation, $P(0^{\otimes Q_{\mathrm{red}}})$ is estimated by repeated measurement as in the SWAP test protocol described above.
The norm $\|\bm{m}_j\|$ and the dimension $D$ are computed classically.
Note that, as with the original SWAP test, this measurement yields only $|\langle U | \tilde{\bm{m}}_j \rangle|^2$; the sign of $S_j$ is determined by $\mathrm{sign}(\sum_l [\bm{m}_j]_l)$, which is computed classically from the known coefficients and basis values.

The merged encoding requires
\begin{equation}
  Q_{\mathrm{red}} = \lceil\log_2(n(d+1))\rceil
  \label{eq:qubits_red}
\end{equation}
qubits per register.

\subsection{Sign recovery}
\label{sec:sign_recovery}

Both the SWAP test and the merged overlap measurement yield only squared overlaps ($|\langle \cdot | \cdot \rangle|^2$), so the sign of $S_j$ is not directly accessible from the quantum measurement.
In our simulation, the sign is computed classically as $\mathrm{sign}(\sum_l [\bm{m}_j]_l)$ from the known coefficients and basis values.
On real quantum hardware, sign recovery would require an additional Hadamard test, which uses one ancilla qubit prepared in $|+\rangle$ and a controlled unitary to extract the real part of the overlap.
This increases the effective qubit count by one ($Q_{\mathrm{red}} + 1$) and adds $O(Q_{\mathrm{red}})$ controlled gates per evaluation.
We emphasize that this is a limitation shared with the original CCQKAN architecture (which also requires sign recovery for the SWAP test), not specific to the merged encoding; both architectures require the same additional ancilla on hardware.

\subsection{Sequential qubit reuse}

The $n$ hidden nodes and the output node are computed sequentially, reusing the same $Q_{\mathrm{red}}$ qubits for each evaluation.
The same merged encoding applies to the second layer: the $n$ second-layer edge activations feeding the output node are merged into a single circuit execution analogous to Eq.~\eqref{eq:merged_vector}, with $x_i$ replaced by $h_i$.
The total qubit count for the full network forward pass is thus $Q_{\mathrm{red}}$ as given in Eq.~\eqref{eq:qubits_red} (or $Q_{\mathrm{red}}+1$ including the sign-recovery ancilla), at the cost of $n+1$ sequential circuit evaluations (one per hidden node plus one for the output).

\subsection{Resource trade-off}
\label{sec:resource}

\begin{table*}[!htb]
\caption{Resource comparison for $[n,n,1]$ CCQKAN. $Q$: qubits; $C$: circuit executions per forward pass. $\Delta Q = Q_{\mathrm{red}} - Q_{\mathrm{seq}}$.}
\label{tab:qubits}
\begin{tabular}{lcrrrrrrr}
Network & $d$ & $Q_{\rm par}$ & $C_{\rm par}$ & $Q_{\rm seq}$ & $C_{\rm seq}$ & $Q_{\rm red}$ & $C_{\rm red}$ & $\Delta Q$ \\
\hline
$[2,2,1]$ & 2 & 12 & 1 & 2 & 6 & 3 & 3 & 1 \\
$[2,2,1]$ & 3 & 12 & 1 & 2 & 6 & 3 & 3 & 1 \\
$[2,2,1]$ & 4 & 18 & 1 & 3 & 6 & 4 & 3 & 1 \\
$[3,3,1]$ & 2 & 24 & 1 & 2 & 12 & 4 & 4 & 2 \\
$[3,3,1]$ & 3 & 24 & 1 & 2 & 12 & 4 & 4 & 2 \\
$[3,3,1]$ & 4 & 36 & 1 & 3 & 12 & 4 & 4 & 1 \\
$[4,4,1]$ & 2 & 40 & 1 & 2 & 20 & 4 & 5 & 2 \\
$[4,4,1]$ & 3 & 40 & 1 & 2 & 20 & 4 & 5 & 2 \\
$[4,4,1]$ & 4 & 60 & 1 & 3 & 20 & 5 & 5 & 2 \\
$[4,4,1]$ & 5 & 60 & 1 & 3 & 20 & 5 & 5 & 2 \\
\end{tabular}
\end{table*}

\FloatBarrier

Table~\ref{tab:qubits} compares the three execution strategies in terms of both qubit count ($Q$) and circuit execution count ($C$) per forward pass.
For a two-layer $[n,n,1]$ network:
\begin{itemize}
  \item Parallel: $C_{\mathrm{par}} = 1$ (all edges evaluated simultaneously),
  \item Sequential: $C_{\mathrm{seq}} = n^2 + n$ (one circuit per edge),
  \item Merged: $C_{\mathrm{red}} = n + 1$ (one circuit per output node).
\end{itemize}
The merged approach reduces circuit executions by a factor of $n$ relative to the sequential baseline:
\begin{equation}
  \frac{C_{\mathrm{red}}}{C_{\mathrm{seq}}} = \frac{n+1}{n^2+n} = \frac{1}{n}.
  \label{eq:circuit_reduction}
\end{equation}
For a $[4,4,1]$ network, this represents a $4\times$ reduction in circuit executions (from 20 to 5).

The qubit overhead of the merged approach relative to the sequential baseline is
\begin{equation}
  \Delta Q = Q_{\mathrm{red}} - Q_{\mathrm{seq}} = \lceil\log_2(n(d+1))\rceil - \lceil\log_2(d+1)\rceil,
  \label{eq:qubit_overhead}
\end{equation}
which grows only as $O(\log n)$.
In all configurations studied, $\Delta Q$ is 1--2 qubits (Table~\ref{tab:qubits}) when both register counts exclude measurement overhead.
However, the two schemes have \emph{asymmetric} measurement overhead: the sequential baseline evaluates each edge by a SWAP test, which requires two $q_e$-qubit registers plus an ancilla ($Q^{\mathrm{SWAP}}_{\mathrm{seq}} = 2q_e + 1$), whereas the merged circuit uses a direct projective measurement on a single register (plus one sign-recovery ancilla).
Under this like-for-like accounting the sign of the comparison reverses: $Q_{\mathrm{red}} + 1 - Q^{\mathrm{SWAP}}_{\mathrm{seq}} \leq 0$ in all 10 configurations (between 0 and $-2$), i.e., the merged encoding reduces circuit executions by a factor of $n$ while using \emph{no more, and typically fewer,} qubits than the sequential SWAP-test baseline.
Whether the merged layout is preferable depends on the relative cost of qubits, circuit executions, and measurement shots on a given hardware platform; we do not attempt to resolve this hardware-dependent question here, but instead focus on characterizing the trainability of the merged encoding.

\begin{figure}[!htb]
  \includegraphics[width=\columnwidth]{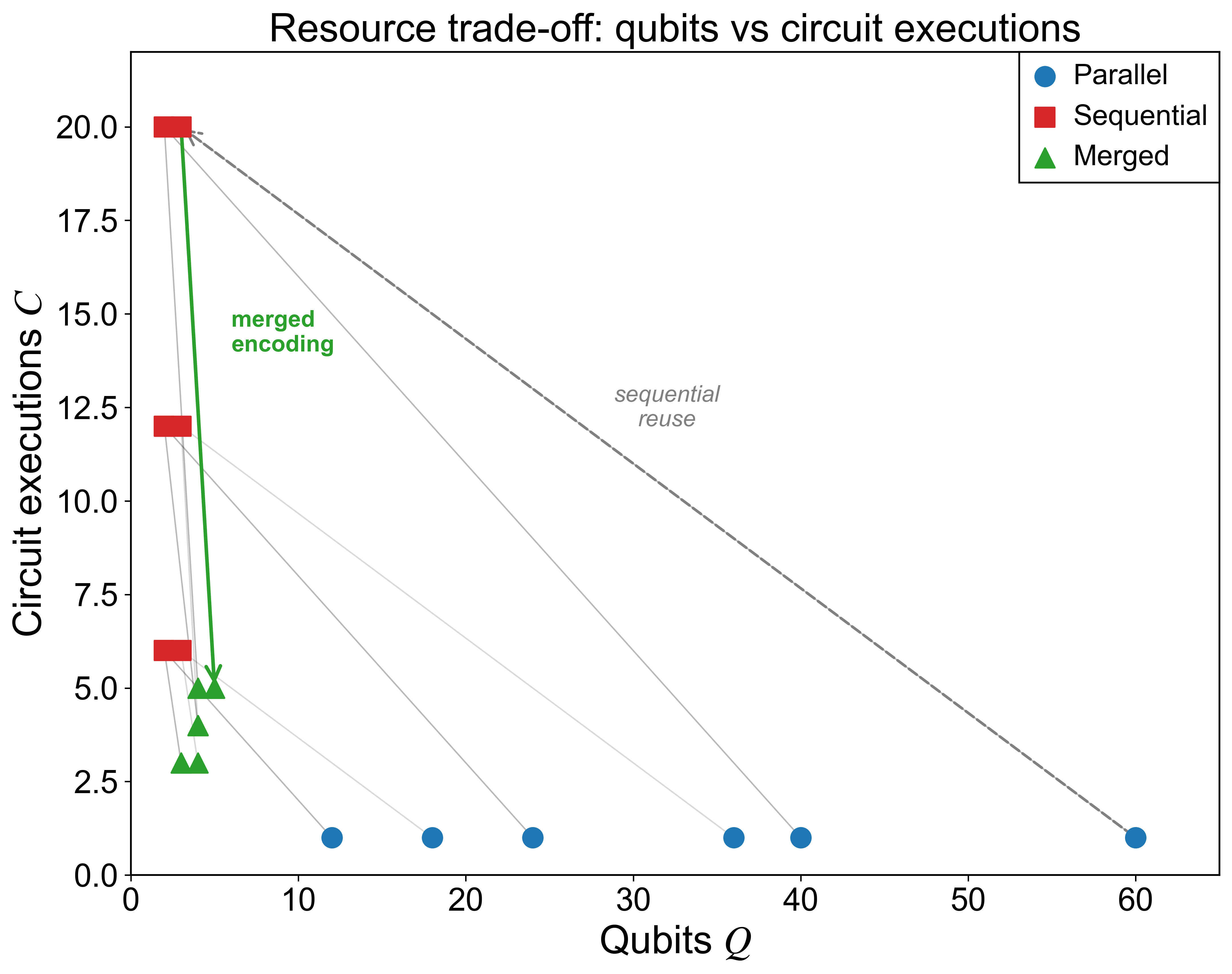}
  \caption{Resource trade-off for all 10 configurations. Each configuration maps to three points: parallel (circle), sequential (square), merged (triangle). The merged approach reduces circuit executions $C$ by a factor of $n$ relative to sequential, at a cost of 1--2 additional qubits.}
  \label{fig:resource_tradeoff}
\end{figure}

Figure~\ref{fig:resource_tradeoff} visualizes this trade-off: each network configuration maps to three points in the $(Q, C)$ plane, illustrating how the merged approach occupies an intermediate position between the parallel (low $C$, high $Q$) and sequential (low $Q$, high $C$) extremes.

The gate complexity of amplitude encoding is an additional factor not captured by qubit count or circuit execution count alone.
Generic amplitude encoding of an arbitrary $D$-dimensional vector requires $O(D)$ CNOT gates~\cite{Shende2006}, which for the merged state ($D = 2^{Q_{\mathrm{red}}}$) is exponential in $Q_{\mathrm{red}}$.
We also note that evaluating sums of inner products is addressed more generally by linear-combination-of-unitaries techniques~\cite{ChildsWiebe2012}; the merged encoding can be viewed as a restricted alternative in which the summands are prepared classically as a single amplitude state rather than block-encoded, trading generality for state-preparation simplicity.
For the sequential baseline, each edge requires amplitude encoding of a $(d+1)$-dimensional vector, i.e., $O(d)$ CNOTs per circuit execution.
The merged encoding encodes an $n(d+1)$-dimensional vector, requiring $O(nd)$ CNOTs per execution.
Thus, while the merged approach reduces circuit executions by a factor of $n$, each execution uses $O(n)$ times more gates, yielding a roughly constant total gate count.
The practical advantage therefore depends on the relative cost of per-execution overhead versus per-gate noise.
Specifically, the merged approach is advantageous when per-execution costs (state preparation, measurement, classical readout, and job scheduling latency) dominate over gate-level noise accumulation.
This scenario arises on cloud-accessible quantum hardware where job submission latency ($\sim$ms per circuit) far exceeds gate times ($\sim$$\mu$s): for a $[4,4,1]$ network, reducing 20 circuit submissions to 5 eliminates 15 round-trip latencies per forward pass, which can dominate total wall-clock time in cloud environments.
Conversely, on hardware where gate errors are the dominant noise source and circuit depth is the limiting factor, the sequential baseline (which uses shallower circuits per execution) may be preferable.

\FloatBarrier

\subsection{Parameter transfer}

When transitioning from the original to the merged architecture, the Chebyshev coefficients $\{c_{ij,k}\}$ can be directly transferred, since both architectures parameterize the same set of univariate functions.
The weight tensor $W^{(1)}_{ijk} = c_{ij,k}$ of the merged first layer is set equal to the trained coefficient of the corresponding edge in the original network; similarly for the second layer and affine output parameters ($\alpha$, $\beta$).
This transfer provides a warm start for subsequent training.
To isolate the effect of the architectural change from the initialization strategy, we study both the case with parameter transfer (Red-T) and the case with independent random initialization (Red-I), where Chebyshev coefficients are drawn from $0.1\,\mathcal{N}(0, 1)$ (the same distribution used for the Original model).

\section{Experimental setup}
\label{sec:experiments}

\subsection{Synthetic function regression}

We evaluate 10 network configurations: $[n,n,1]$ for $n \in \{2,3,4\}$ with Chebyshev degree $d \in \{2,3,4\}$, plus $d=5$ for $n=4$ (Table~\ref{tab:qubits}).
The target functions are
\begin{align}
  f_2(\bm{x}) &= \sin(x_1+x_2), \nonumber\\
  f_3(\bm{x}) &= \sin(x_1+x_2)+\cos(x_3), \nonumber\\
  f_4(\bm{x}) &= \sin(x_1+x_2)+\cos(x_3 x_4).
  \label{eq:target_functions}
\end{align}
Training data consists of 30 points sampled uniformly from $[-1,1]^n$, with targets normalized to $[-2,2]$.
The small training set is chosen to keep the computational cost of the 16-seed $\times$ 10-configuration $\times$ 3-condition grid tractable; all comparisons use the training loss (not a held-out test loss), so the relevant question is whether the two architectures optimize the same objective equally well, rather than whether either generalizes.

The loss function used throughout all synthetic experiments is the mean squared error (MSE):
\begin{equation}
  \mathcal{L}(\bm{\theta}) = \frac{1}{N}\sum_{i=1}^{N}\left(\hat{y}_i(\bm{\theta}) - y_i\right)^2,
  \label{eq:mse_loss}
\end{equation}
where $\hat{y}_i(\bm{\theta})$ is the network output for sample $i$, $y_i$ is the target value, $N=30$ is the training set size, and $\bm{\theta}$ denotes the full parameter vector (Chebyshev coefficients, output scale $\alpha$ and shift $\beta$).

Training uses the Adam optimizer~\cite{Kingma2015Adam} with learning rate $\eta = 0.05$ (larger than the Adam default to compensate for the limited 20-step training budget; the reference CCQKAN implementation~\cite{GonzalezOtero2025} uses a step size of 0.3 with the same 20-epoch budget in its classification experiments) and hyperparameters $\beta_1 = 0.9$, $\beta_2 = 0.999$, $\epsilon = 10^{-8}$.
Gradients are computed via central finite differences:
\begin{equation}
  \frac{\partial \mathcal{L}}{\partial \theta_j} \approx \frac{\mathcal{L}(\bm{\theta} + \epsilon \bm{e}_j) - \mathcal{L}(\bm{\theta} - \epsilon \bm{e}_j)}{2\epsilon},
  \label{eq:numerical_gradient}
\end{equation}
with $\epsilon = 10^{-5}$ (a standard choice for double-precision finite differences, balancing truncation and round-off errors) and $\bm{e}_j$ the $j$-th unit vector.
Each model is trained for 20 steps.

\subsection{Simulation conditions}
\label{sec:conditions}

Three simulation conditions are tested:
\begin{enumerate}
  \item \textbf{Ideal}: exact statevector inner products computed as $\langle \hat{\bm{c}} | \hat{\bm{T}} \rangle = \mathrm{Re}[\langle \hat{\bm{c}}^* \cdot \hat{\bm{T}} \rangle]$ (no stochastic noise).
  For the merged circuit, $\langle U | \tilde{\bm{m}}_j \rangle$ is computed analogously.
  \item \textbf{Shot noise}: The SWAP test [Eq.~\eqref{eq:swap_test}] is simulated with $N_{\mathrm{shots}} = 1000$ measurement shots.
  The ancilla measurement is sampled from a binomial distribution: $n_0 \sim \mathrm{Binom}(N_{\mathrm{shots}}, P(0))$, and the overlap squared is estimated as $\widehat{|\langle \cdot | \cdot \rangle|^2} = \max(2 n_0 / N_{\mathrm{shots}} - 1, 0)$ (the clamping to zero introduces a small positive bias when the true overlap is near zero, which affects both architectures equally).
  The sign of the overlap is taken from the exact calculation; on real hardware, the sign is not directly accessible from the SWAP test, which outputs only $|\langle a | b \rangle|^2$.
  Sign recovery on hardware would require additional circuitry (e.g., a Hadamard test); this is a limitation of our simulation protocol.
  Each inner product evaluation during training requires $N_{\mathrm{shots}}$ measurements; the total number of circuit executions per gradient step is $2|\bm{\theta}| \cdot N \cdot C$, where $|\bm{\theta}|$ is the number of parameters, $N$ is the training set size, and $C$ is the circuit execution count per forward pass ($C_{\mathrm{seq}} = n^2+n$ for Sequential, $C_{\mathrm{red}} = n+1$ for Merged).
For example, a $[4,4,1]$ network with $d=5$ has $|\bm{\theta}| = 122$ parameters and $N = 30$: the merged approach requires $2 \times 122 \times 30 \times 5 = 36{,}600$ circuit executions per gradient step versus $146{,}400$ for sequential, each multiplied by $N_{\mathrm{shots}}$ measurements.
  \item \textbf{Shot + device noise}: As condition (2), with depolarizing noise $p=0.01$ applied to both state preparations.
  The noisy density matrix is $\rho_{\mathrm{noisy}} = (1-p)|\psi\rangle\langle\psi| + p\,I/D$, where $D = 2^{Q}$ is the Hilbert space dimension and $Q$ is the qubit count of the register.
  Because the SWAP-test outcome probability depends on the states only through $\mathrm{Tr}(\rho_a \rho_b)$, depolarizing noise is applied \emph{exactly} in closed form: $\mathrm{Tr}(\rho_a \rho_b) = (1-p)^2 |\langle a | b \rangle|^2 + 2p(1-p)/D + p^2/D$, and the ancilla probability $P(0) = [1 + \mathrm{Tr}(\rho_a \rho_b)]/2$ is then sampled binomially as in condition (2). No pure-state approximation is involved.
  We caution that a pure-state shortcut---approximating the noisy state by the dominant eigenvector of $\rho_{\mathrm{noisy}}$---cannot represent this channel: that eigenvector is exactly $|\psi\rangle$ for every $p<1$, so such an approximation is a null operation. All condition-(3) results use the exact mixed-state formula above.
This is a simplification; the conclusions drawn from this noise condition should be interpreted as qualitative rather than quantitative predictions for real hardware behavior.
\end{enumerate}

For each condition, three models are compared: Original, Reduced with parameter transfer (Red-T), and Reduced with independent initialization (Red-I).
In the Original model, each of the $(n^2+n)$ edges computes its inner product independently using $q_e$ qubits.
In the Reduced model, each of the $(n+1)$ output nodes computes its merged inner product using $Q_{\mathrm{red}}$ qubits.
Both models use the same loss function [Eq.~\eqref{eq:mse_loss}], optimizer, training protocol, and---crucially---the same number of trainable parameters (the Chebyshev coefficients $\{c_{ij,k}\}$, scale $\alpha$, and shift $\beta$); the merged encoding changes only how the forward pass is evaluated, not the parameter space.

\subsection{Statistical testing}
\label{sec:stat_testing}

Loss curves with standard deviation are computed over 10 seeds (seeds 0--9).
For statistical significance testing, we perform 16 independent runs (seeds 0--15) under each condition, where the seed controls both parameter initialization and the random number generator for shot-noise sampling (but the training data remain fixed).
For each seed, the Original and Red-I models receive \emph{different} random initializations (each drawn independently from $0.1\,\mathcal{N}(0,1)$), while the Red-T model inherits its initialization from the trained Original; the Wilcoxon pairing is by seed, meaning each pair shares the same training data but not the same initial parameters or shot-noise realizations (the global random stream is consumed by the Original's training before the Red-I model is constructed).
We apply the Wilcoxon signed-rank test~\cite{Wilcoxon1945} to paired final MSE losses.
Significance levels are $^{***}$ ($p<0.001$), $^{**}$ ($p<0.01$), $^{*}$ ($p<0.05$), and n.s.\ ($p \geq 0.05$).

We note that the Wilcoxon test assesses whether the paired differences have a symmetric distribution around zero (testing $H_0$: the median difference is zero), so a non-significant result does not \emph{prove} equivalence---it only indicates failure to detect a difference at the given sample size.
Formally establishing equivalence would require a two one-sided tests (TOST) procedure with a pre-specified equivalence margin~$\delta$.
We do not perform TOST because the choice of $\delta$ is arbitrary for a new architecture without an established performance baseline.
Instead, we supplement the Wilcoxon results with effect sizes (rank-biserial correlation $r$, computed as $r = 1 - 2T/[n(n+1)/2]$ where $T$ is the Wilcoxon statistic).
For the Original vs.\ Red-I comparison under ideal conditions, the mean $|r|$ across all 10 configurations is 0.25 (range 0.015--0.81; the maximum corresponds to the single significant case $[2,2,1]\ d\!=\!4$).
Excluding that outlier, the mean $|r|$ is 0.19, indicating a small effect.
The corresponding Cohen's $d$ values range from $-0.49$ to $+0.29$ (9 of 10 configurations), confirming that any difference between the two architectures is small relative to inter-seed variability.
Under shot-noise and shot-plus-noise conditions, the mean $|r|$ values are 0.22 and 0.33, respectively, with no systematic direction of effect (some configurations favor Original, others favor Red-I).

We acknowledge a limitation of statistical power: with $n=16$ paired samples, the Wilcoxon test has approximately 80\% power to detect a large effect ($r \approx 0.7$) at $\alpha = 0.05$, but only $\sim$45\% power for a medium effect ($r \approx 0.5$).
The non-significant results should therefore be interpreted conservatively: they indicate that the merged encoding does not produce a \emph{large} degradation in trainability, but a moderate effect cannot be ruled out at this sample size.
The consistency of the non-significant results across 30 comparisons (10 configurations $\times$ 3 conditions) strengthens the evidence beyond what any single test provides, though these comparisons are not fully independent (the same 10 configurations appear under each condition, and noisy conditions share the same underlying parameters).
We do not apply a Bonferroni or similar correction because each comparison addresses the same directional hypothesis (merged encoding does not degrade performance) rather than testing for any difference among many hypotheses; however, we note that the two significant results ($p < 0.05$) out of 30 are consistent with the nominal Type~I error rate.

\subsection{Digit classification}

To validate on real data, we perform two classification tasks on the scikit-learn~\cite{Pedregosa2011sklearn} $8\times8$ digit dataset.
Input dimensionality is reduced from 64 to $n$ via principal component analysis (PCA), and inputs are normalized to $[-1,1]$ via min-max scaling.
The loss function is the same MSE as in the synthetic experiments:
\begin{equation}
  \mathcal{L}_{\mathrm{MNIST}} = \frac{1}{N}\sum_{i=1}^{N}\left(\hat{y}_i - y_i\right)^2.
  \label{eq:mnist_loss}
\end{equation}
Both tasks use 100 training samples and 50 common test samples with 10 random training splits (fewer than the 16 seeds used for synthetic experiments, due to the higher per-run cost of training 10 OvA classifiers), under ideal statevector simulation.

\textbf{Binary classification (0 vs 1).}
Labels are $\{-1, +1\}$, and classification is by $\mathrm{sign}(\hat{y})$.
Configurations are $[n,n,1]$ for $n \in \{2,3,4,5,6\}$ with $d=2$.
All three models (Original, Red-T, Red-I) are compared.

\textbf{10-class classification (one-vs-all).}
To test on a more challenging task, we train 10 independent binary $[n,n,1]$ classifiers per model type, one for each digit class $c \in \{0,\ldots,9\}$.
For classifier $c$, labels are $y_i = +1$ if the digit is $c$ and $y_i = -1$ otherwise.
The predicted class is $\hat{c} = \arg\max_{c} \hat{y}_c$, where $\hat{y}_c$ is the raw output (before sign) of classifier $c$.
Configurations are $[n,n,1]$ for $n \in \{2,3,4,5,6\}$ with $d=3$ (chosen to provide more expressive per-edge activation functions than $d=2$ while remaining computationally tractable for the 10-classifier setup).
Only Original and Red-I (independently initialized) are compared.
Red-T is excluded because the 10-class experiment trains 10 separate OvA classifiers per configuration, each requiring its own pre-trained Original model; this would triple the already substantial computational cost ($\sim$74 hours for the current setup) without addressing the paper's central question of whether the merged architecture is trainable from scratch.

\section{Results}
\label{sec:results}

\subsection{Ideal conditions}

\begin{figure*}[!htb]
  \includegraphics[width=\textwidth]{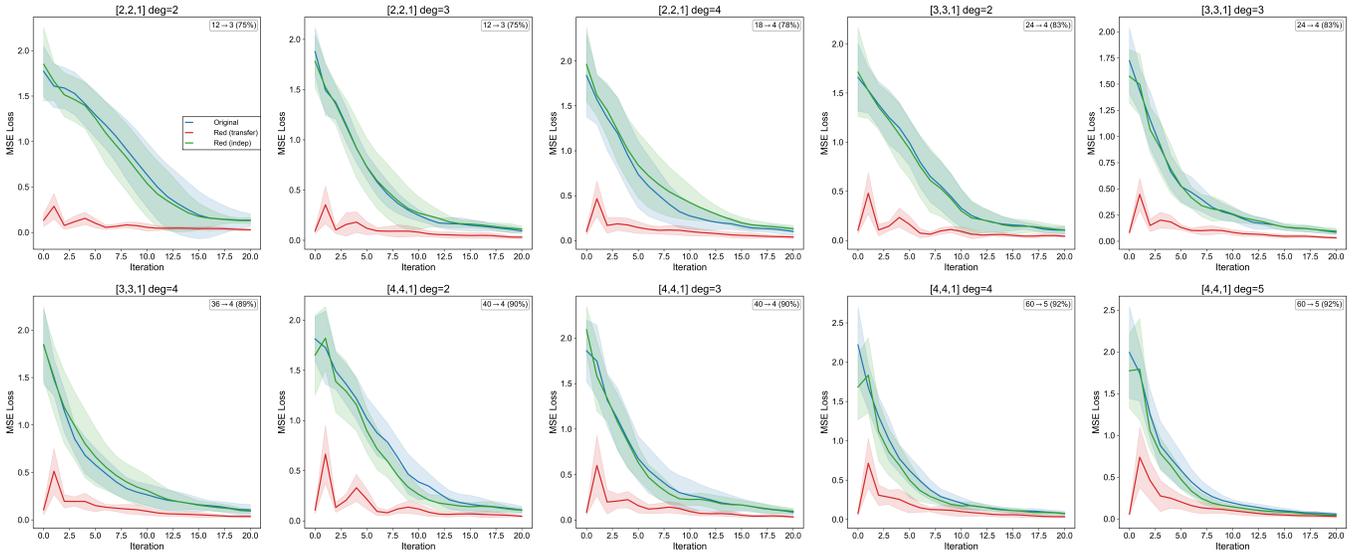}
  \caption{Training loss curves under ideal conditions for all 10 configurations ($[n,n,1]$, degrees as labeled). Solid lines: mean over seeds 0--9; shaded regions: $\pm 1$ standard deviation. Blue: Original, orange: Red-T (parameter transfer), green: Red-I (independent initialization).}
  \label{fig:ideal_loss}
\end{figure*}

Figure~\ref{fig:ideal_loss} shows the loss curves under ideal conditions for all 10 configurations.
Red-T converges faster and to lower loss than both Original and Red-I, reflecting the warm-start advantage of parameter transfer.
Red-I tracks Original closely throughout training.

As shown in Table~\ref{tab:ideal}, Red-T consistently achieves the lowest final MSE across all 10 configurations in ideal conditions, with improvements of 48--78\% over Original.
We emphasize that this comparison is budget-confounded: Red-T receives 20 additional Adam steps on top of the 20 already invested in training the Original it inherits from, for an effective 40-step budget versus 20.
A budget-matched control (companion package, 10 configurations $\times$ 16 seeds) confirms this reading: an Original trained for 40 steps and a Red-T (20+20) reach final losses identical to four decimal places in every configuration (all Wilcoxon $p > 0.2$), as dictated by the shared loss landscape.
The Red-T advantage over the 20-step Original is therefore an optimization-budget effect, not an architectural one.
Red-I achieves losses comparable to Original.
As discussed in Sec.~\ref{sec:expressiveness}, this is expected under ideal conditions since both architectures compute the same function and follow the same optimization trajectory from the same loss landscape; the ideal-condition comparison thus serves primarily as a consistency check.
The Wilcoxon test (Table~\ref{tab:sig_ideal}) confirms that Original vs.\ Red-T is significant at $p < 0.003$ in all 10 configurations, while Original vs.\ Red-I is not significant in 9 of 10 cases.
Red-T vs.\ Red-I is significant at $p < 0.001$ in 9 of 10 cases, confirming that the Red-T advantage is attributable entirely to the warm-start initialization rather than any property of the merged architecture itself.

\begin{figure*}[!htb]
  \includegraphics[width=\textwidth]{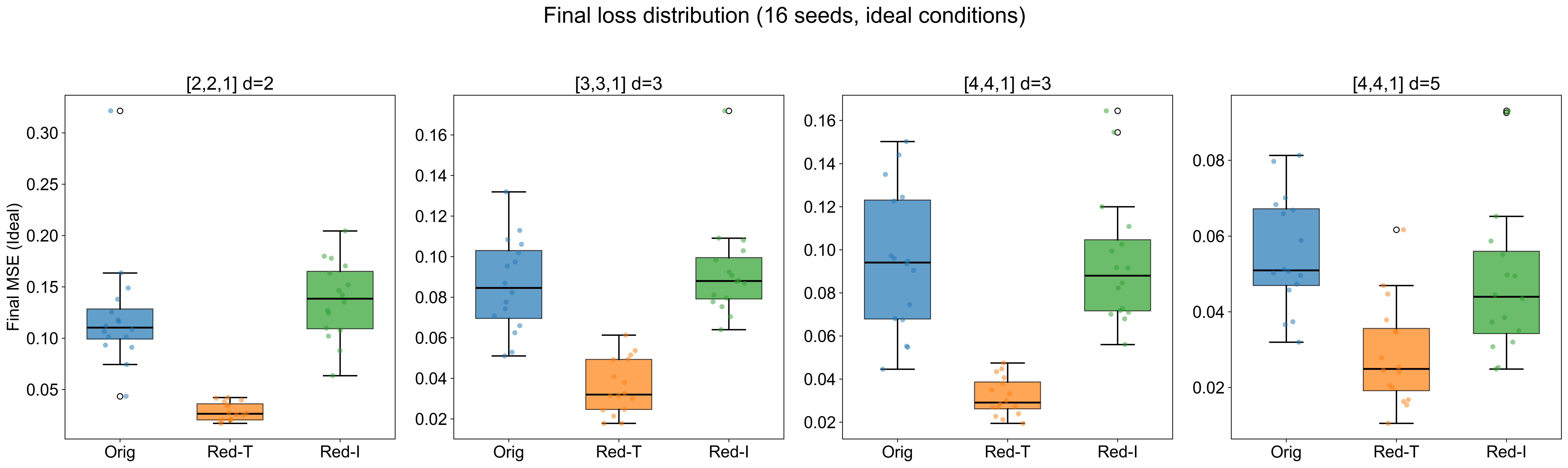}
  \caption{Final MSE loss distributions under ideal conditions (16 seeds) for four representative configurations. Box: interquartile range; line: median; dots: individual seeds. Blue: Original, orange: Red-T, green: Red-I. The overlap between Original and Red-I distributions is consistent with the non-significant Wilcoxon results.}
  \label{fig:boxplot_ideal}
\end{figure*}

Figure~\ref{fig:boxplot_ideal} shows the distribution of final MSE losses across 16 seeds for four representative configurations.
The box plots reveal that Red-T has both lower median and smaller interquartile range than Original and Red-I, while the distributions of Original and Red-I overlap substantially, consistent with the non-significant Wilcoxon results.

\begin{table*}[!htb]
\caption{Final MSE loss under ideal conditions. Left three columns: mean $\pm$ std over seeds 0--9 (used for loss curves in Fig.~\ref{fig:ideal_loss}). Right three columns: Wilcoxon signed-rank test over seeds 0--15 (the additional 6 seeds provide greater statistical power). Bold indicates the lowest mean loss for each configuration. O = Original, T = Red-T, I = Red-I.}
\label{tab:ideal}
\begin{tabular}{lcccccc}
Network & Original & Red-T & Red-I & O vs T & O vs I & T vs I \\
\hline
$[2,2,1]\ d\!=\!2$ & $.135 \pm .070$ & $\mathbf{.030 \pm .009}$ & $.130 \pm .035$ & $^{***}$ & n.s. & $^{***}$ \\
$[2,2,1]\ d\!=\!3$ & $.093 \pm .026$ & $\mathbf{.031 \pm .015}$ & $.111 \pm .029$ & $^{***}$ & n.s. & $^{***}$ \\
$[2,2,1]\ d\!=\!4$ & $.102 \pm .037$ & $\mathbf{.039 \pm .019}$ & $.132 \pm .037$ & $^{***}$ & $^{**}$ & $^{***}$ \\
$[3,3,1]\ d\!=\!2$ & $.105 \pm .037$ & $\mathbf{.043 \pm .010}$ & $.106 \pm .042$ & $^{***}$ & n.s. & $^{***}$ \\
$[3,3,1]\ d\!=\!3$ & $.085 \pm .024$ & $\mathbf{.033 \pm .009}$ & $.094 \pm .029$ & $^{***}$ & n.s. & $^{***}$ \\
$[3,3,1]\ d\!=\!4$ & $.106 \pm .056$ & $\mathbf{.036 \pm .014}$ & $.092 \pm .023$ & $^{***}$ & n.s. & $^{***}$ \\
$[4,4,1]\ d\!=\!2$ & $.108 \pm .036$ & $\mathbf{.046 \pm .010}$ & $.105 \pm .023$ & $^{***}$ & n.s. & $^{***}$ \\
$[4,4,1]\ d\!=\!3$ & $.086 \pm .028$ & $\mathbf{.034 \pm .009}$ & $.096 \pm .035$ & $^{***}$ & n.s. & $^{***}$ \\
$[4,4,1]\ d\!=\!4$ & $.077 \pm .031$ & $\mathbf{.034 \pm .013}$ & $.073 \pm .012$ & $^{**}$  & n.s. & $^{***}$ \\
$[4,4,1]\ d\!=\!5$ & $.058 \pm .015$ & $\mathbf{.030 \pm .016}$ & $.042 \pm .011$ & $^{***}$ & n.s. & $^{*}$ \\
\end{tabular}
\label{tab:sig_ideal}
\end{table*}

\FloatBarrier

\FloatBarrier

\subsection{Shot noise conditions}

Under 1000-shot measurement noise, the MSE increases substantially for all models (from $\sim$0.03--0.13 to $\sim$2.1--3.9), reflecting the dominant effect of statistical estimation error in the SWAP test.
Given that the target range is $[-2,2]$ (MSE $\leq 4$ for random predictions), these loss values indicate that shot noise largely overwhelms the learning signal at this shot budget, limiting the discriminative power of this condition.
The dominant cause is the gradient estimator rather than the circuits: central finite differences with $\epsilon = 10^{-5}$ amplify the shot-level standard error of the loss by $1/(2\epsilon) = 5 \times 10^{4}$, so the optimizer receives essentially random directions and training frequently ends above its initial loss.

\begin{figure*}[!htb]
  \includegraphics[width=\textwidth]{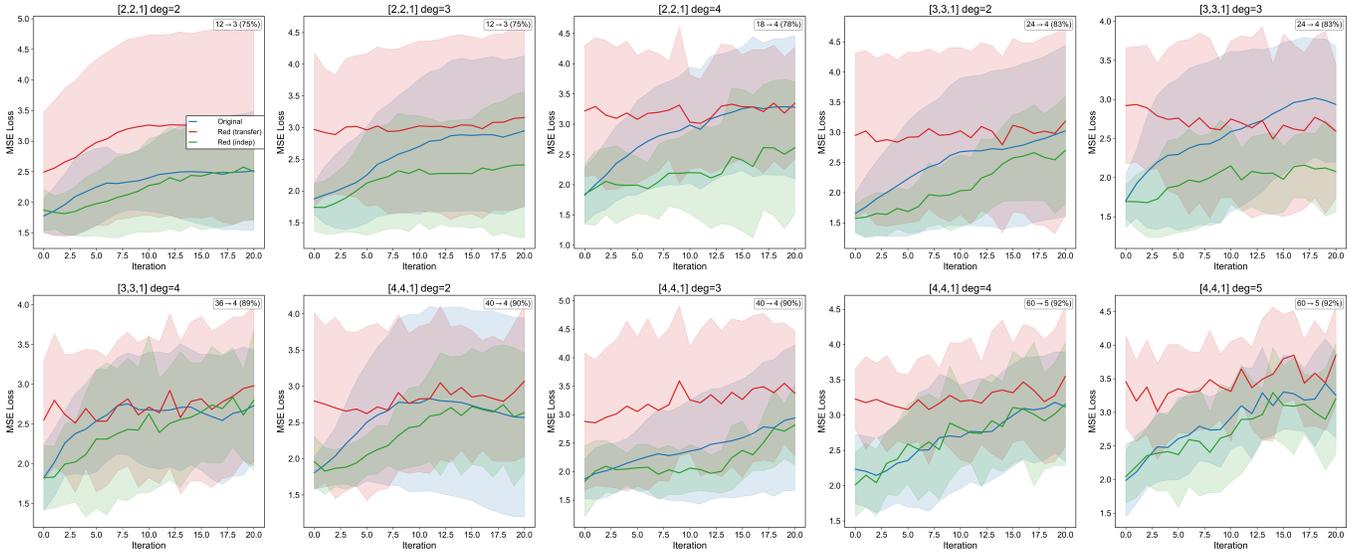}
  \caption{Training loss curves under 1000-shot measurement noise. Colors as in Fig.~\ref{fig:ideal_loss}. Shot noise dominates the loss, compressing the differences between models.}
  \label{fig:shots_loss}
\end{figure*}

To measure trainability with a hardware-standard estimator, we reran the finite-shot condition using SPSA (two loss evaluations per step, standard decaying gains) at a total measurement budget matched to the 20-step finite-difference protocol, i.e.\ $20|\bm{\theta}|$ SPSA steps (companion package, 10 configurations $\times$ 16 seeds, 1000 shots).
Under SPSA, training converges in all 320 runs (final loss below initial in 16/16 seeds for every configuration and both architectures), and the condition becomes genuinely discriminative.
The outcome is architecture-dependent: at $n = 2$ the two circuits are indistinguishable ($p > 0.25$), while for $n \geq 3$ the merged circuit reaches systematically higher final loss (significant in 6 of 10 configurations, e.g.\ $0.0119$ vs.\ $0.0240$ at $[4,4,1]$ $d{=}4$, $p < 10^{-4}$; aggregate configuration-level $t$-test $p = 0.017$, Wilcoxon $p = 0.020$, merged worse in 8 of 10).
This deficit, which grows with $n$, is consistent with the $\|\bm{m}_j\|\sqrt{D}$ amplification of measurement noise in the merged estimator [Eq.~\eqref{eq:Sj_recovery}] and independently corroborates the aggregate result of the 10{,}000-shot supplementary experiment (Sec.~\ref{sec:limitations}).
We therefore qualify the trade-off: the $n$-fold reduction in circuit executions is obtained at the cost of a mild, $n$-dependent degradation of finite-shot trainability at fixed per-execution shot count.
The results are summarized in Table~\ref{tab:sig_summary} and Fig.~\ref{fig:shots_loss}.
Red-I achieves performance statistically indistinguishable from Original in all 10 configurations.
Red-T performs slightly worse than Original, with 4 of 10 configurations reaching significance ($p < 0.05$), suggesting that parameters optimized under ideal conditions do not transfer well to the noisy regime.

\FloatBarrier

\subsection{Shot noise + depolarizing noise}

\begin{figure*}[!htb]
  \includegraphics[width=\textwidth]{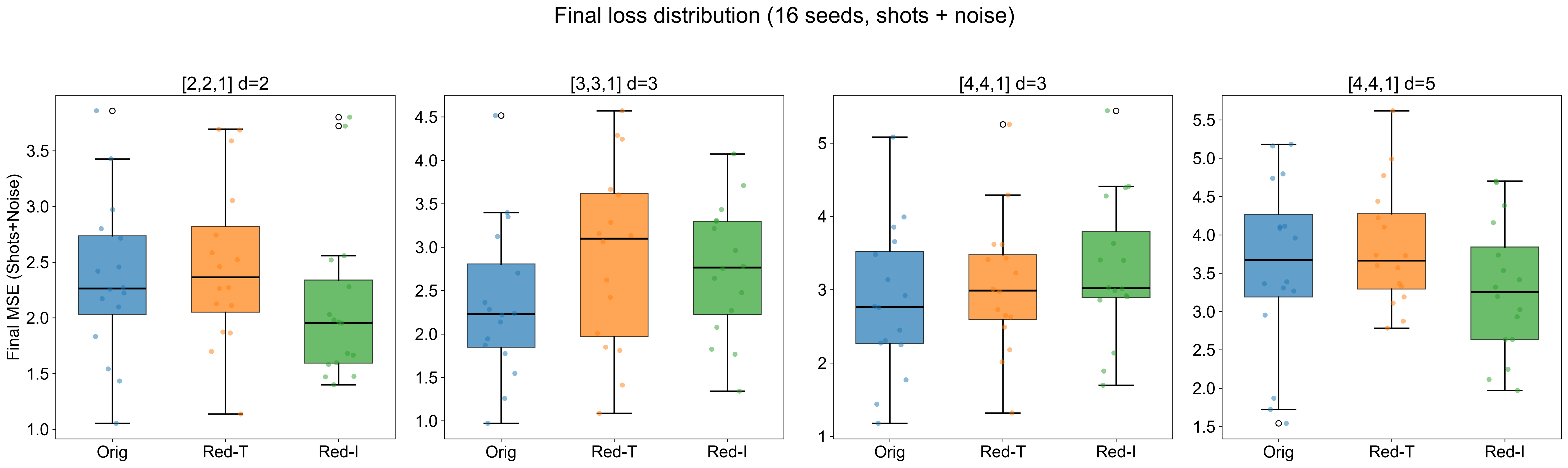}
  \caption{Final MSE loss distributions under shot noise plus depolarizing noise ($p=0.01$) for four representative configurations (16 seeds). Colors as in Fig.~\ref{fig:boxplot_ideal}. All three model distributions overlap substantially, consistent with the non-significant Wilcoxon results.}
  \label{fig:boxplot_noise}
\end{figure*}

\begin{figure*}[!htb]
  \includegraphics[width=\textwidth]{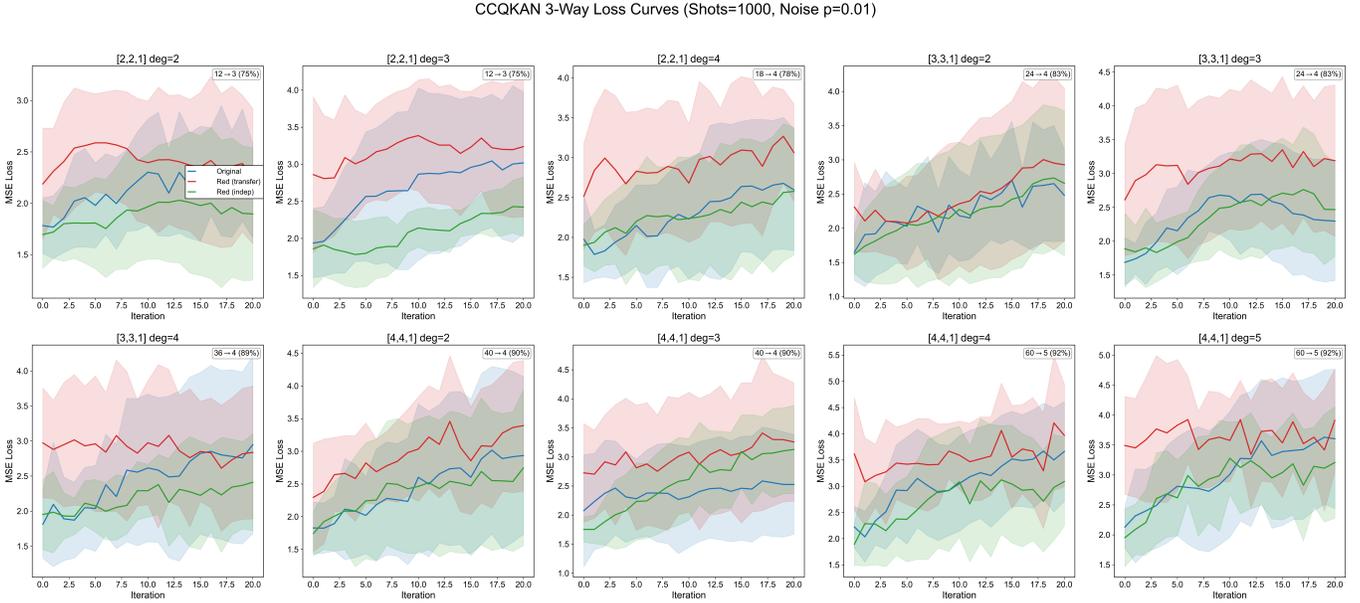}
  \caption{Training loss curves under 1000-shot measurement noise plus depolarizing noise ($p=0.01$). Colors as in Fig.~\ref{fig:ideal_loss}. All three models converge to similar loss values.}
  \label{fig:noise_loss}
\end{figure*}

Adding depolarizing noise ($p=0.01$) further equalizes performance: all three pairwise comparisons are non-significant in 9/10 or 10/10 configurations (Table~\ref{tab:sig_summary}, Figs.~\ref{fig:noise_loss}--\ref{fig:boxplot_noise}), consistent with the noise-induced flattening of the optimization landscape established for variational circuits~\cite{Wang2021noise,Stilck2021}.
An independent rerun using the exact mixed-state noise formula above (companion package, 10 configurations $\times$ 16 seeds, 1000 shots, $p = 0.01$) likewise finds no systematic difference between Original and Red-I: the merged circuit is worse on average in 6 of 10 configurations with an aggregate configuration-level $p \approx 0.43$, and the three per-configuration significant cases split in direction (two favoring Original, one favoring the merged circuit).

\FloatBarrier

\begin{table*}[!htb]
\caption{Significance summary across all conditions (Wilcoxon signed-rank test, 16 seeds). Each entry shows the number of configurations (out of 10) reaching each significance level. ``Lower loss'' indicates which model has the lower mean loss among the significant cases.}
\label{tab:sig_summary}
\begin{tabular}{llccccl}
Comparison & Condition & $p<0.001$ & $p<0.01$ & $p<0.05$ & $p\geq0.05$ & Lower loss \\
\hline
\multirow{3}{*}{Original vs Red-T} & Ideal     & 9 & 1 & 0 & 0  & Red-T \\
                                    & Shots     & 0 & 2 & 2 & 6  & Original \\
                                    & Shots+Noise & 0 & 0 & 0 & 10 & -- \\
\hline
\multirow{3}{*}{Original vs Red-I} & Ideal     & 0 & 1 & 0 & 9  & -- \\
                                    & Shots     & 0 & 0 & 0 & 10 & -- \\
                                    & Shots+Noise & 0 & 0 & 1 & 9  & -- \\
\hline
\multirow{3}{*}{Red-T vs Red-I}    & Ideal     & 9 & 0 & 1 & 0  & Red-T \\
                                    & Shots     & 0 & 0 & 2 & 8  & -- \\
                                    & Shots+Noise & 0 & 0 & 1 & 9  & -- \\
\end{tabular}
\end{table*}

\FloatBarrier

\subsection{Statistical significance overview}

\begin{figure*}[!htb]
  \includegraphics[width=\textwidth]{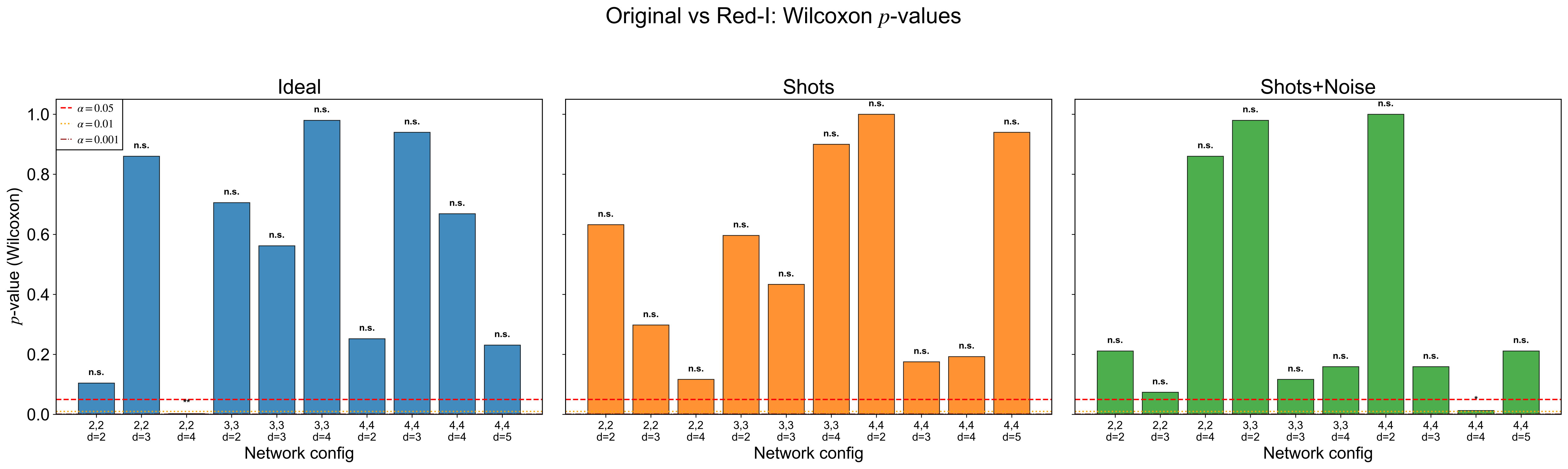}
  \caption{Wilcoxon $p$-values for the Original vs.\ Red-I comparison across all 10 configurations and three conditions (16 seeds). Dashed lines mark significance thresholds $\alpha = 0.05$, $0.01$, $0.001$. Labels above bars indicate significance levels. Nearly all comparisons are non-significant.}
  \label{fig:pvalue_bar}
\end{figure*}

Figure~\ref{fig:pvalue_bar} shows the Wilcoxon $p$-values for the Original vs.\ Red-I comparison across all conditions.
Under all three conditions, 28 of 30 configuration--condition pairs are non-significant at $\alpha = 0.05$ (Fig.~\ref{fig:pvalue_bar}).
The two significant cases (2 out of 30) are consistent with the expected Type~I error rate ($30 \times 0.05 = 1.5$), consistent with the merged encoding preserving the optimization dynamics of the original circuit.

\begin{figure*}[!htb]
  \includegraphics[width=\textwidth]{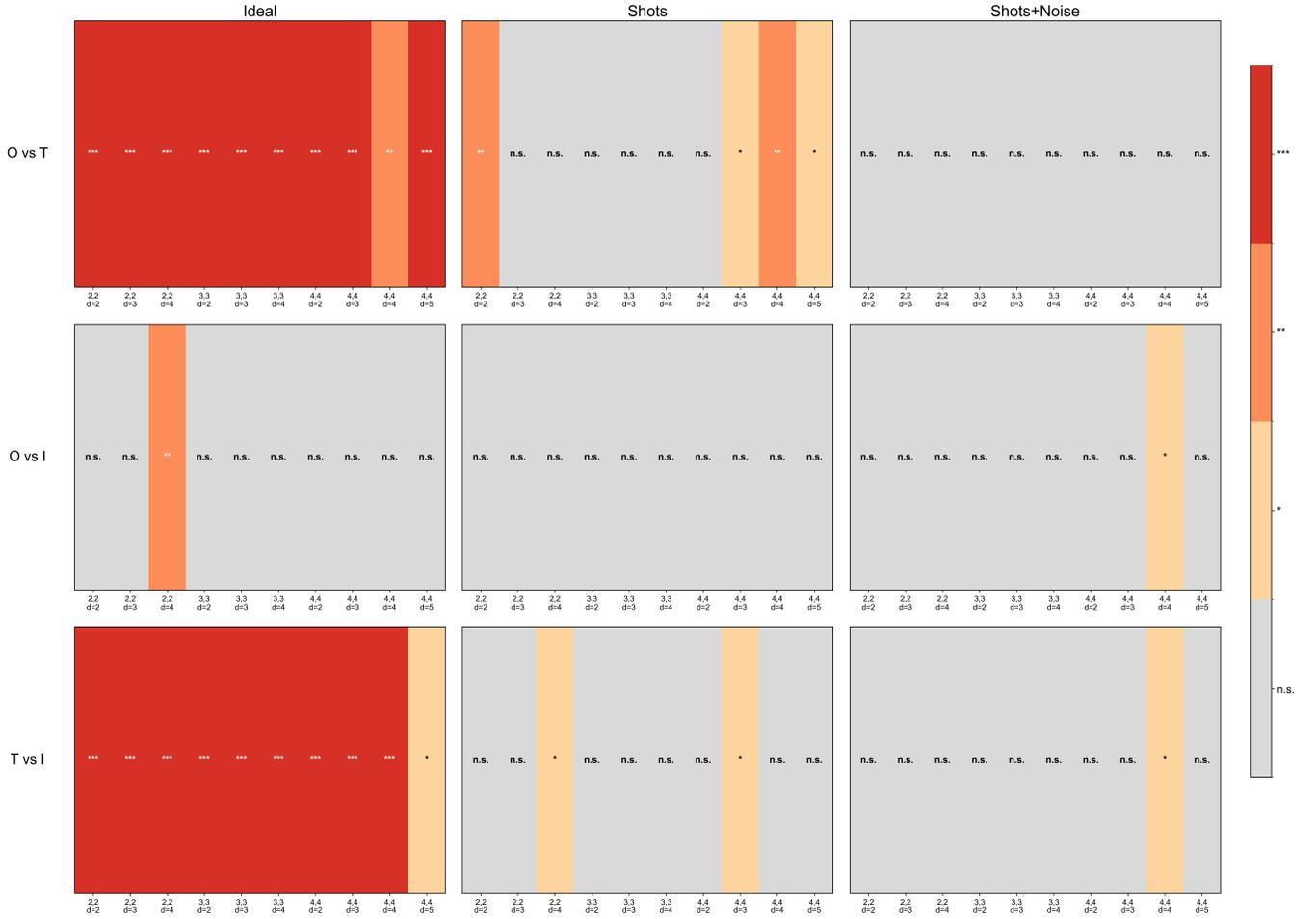}
  \caption{Full significance heatmap (Wilcoxon, 16 seeds). Rows: three pairwise comparisons (O vs T, O vs I, T vs I). Columns: three conditions (Ideal, Shots, Shots+Noise). Each cell shows the significance level for one of 10 configurations. Gray: $p \geq 0.05$; progressively darker shading indicates $^{*}$, $^{**}$, $^{***}$.}
  \label{fig:sig_heatmap}
\end{figure*}

Figure~\ref{fig:sig_heatmap} provides a comprehensive overview of all nine pairwise comparisons (3 pairs $\times$ 3 conditions).
The Original vs.\ Red-T row (top) shows uniformly significant results under ideal conditions, transitioning to mostly non-significant under noise.
The Original vs.\ Red-I row (middle) is almost entirely non-significant across all conditions.

\FloatBarrier

\subsection{Digit classification}

As a supplementary consistency check, we apply all three models to binary digit classification (0 vs 1) on the $8\times8$ scikit-learn digits dataset under ideal conditions.
All three models achieve test accuracy $\geq 99.8\%$ across all network sizes ($n=2$ to $6$); the task is nearly linearly separable after PCA and has no discriminative power for distinguishing architectures.
The MSE loss follows the same pattern as the synthetic experiments: Red-T achieves the lowest loss, while Red-I and Original are comparable (see Table~\ref{tab:mnist}).

\begin{table}[!htb]
\caption{MNIST 0-vs-1: final MSE loss (mean $\pm$ std, 10 splits, $d=2$, ideal). All test accuracies $\geq 99.8\%$.}
\label{tab:mnist}
\begin{tabular}{lccc}
 & Original & Red-T & Red-I \\
\hline
$[2,2,1]$ & $0.050 \pm 0.024$ & $\mathbf{0.004 \pm 0.002}$ & $0.040 \pm 0.009$ \\
$[3,3,1]$ & $0.038 \pm 0.012$ & $\mathbf{0.009 \pm 0.004}$ & $0.040 \pm 0.020$ \\
$[4,4,1]$ & $0.030 \pm 0.010$ & $\mathbf{0.007 \pm 0.004}$ & $0.029 \pm 0.013$ \\
$[5,5,1]$ & $0.025 \pm 0.006$ & $\mathbf{0.014 \pm 0.004}$ & $0.031 \pm 0.008$ \\
$[6,6,1]$ & $0.032 \pm 0.010$ & $\mathbf{0.013 \pm 0.006}$ & $0.030 \pm 0.013$ \\
\end{tabular}
\end{table}

\FloatBarrier

\subsection{10-class digit classification}
\label{sec:10class}

Table~\ref{tab:mc_mnist} summarizes the 10-class one-vs-all (OvA) classification results.
Unlike the binary task, where all models achieved near-perfect accuracy, the 10-class task produces test accuracies ranging from 53\% to 78\%, well above the 10\% chance level but far from perfect, providing genuine discriminative power for comparing architectures.
For context, classical linear baselines (logistic regression and linear SVM) on the same PCA-reduced features and training protocol achieve 43--77\% test accuracy, indicating that CCQKAN's performance is comparable to---and at $n=6$ slightly below---classical linear models on this low-dimensional task; we do not claim any quantum advantage from these results.

Accuracy increases monotonically with $n$ up to $n=5$, then decreases slightly at $n=6$.
Crucially, this non-monotonicity is observed in \emph{both} Original ($78.0\% \to 74.2\%$) and Red-I ($77.8\% \to 74.6\%$), indicating that it is an architecture-general phenomenon unrelated to the merged encoding.
The likely cause is insufficient training data relative to model capacity: the $[6,6,1]$ network has 168 Chebyshev coefficients per OvA classifier at $d=3$ (vs.\ 120 for $[5,5,1]$), while the training set contains only 100 samples.
The train accuracy at $n=6$ (87.6\%) is comparable to $n=5$ (89.8\%), suggesting that the additional PCA variance (59.4\% vs.\ 54.5\%) does not compensate for the increased parameter count in this data-limited regime.
The maximum absolute difference between Original and Red-I test accuracy is 2.2 percentage points ($[2,2,1]$ and $[4,4,1]$), and the Wilcoxon signed-rank test confirms no significant difference in any of the five configurations (Table~\ref{tab:mc_wilcoxon}).

\begin{figure*}[!htb]
  \includegraphics[width=\textwidth]{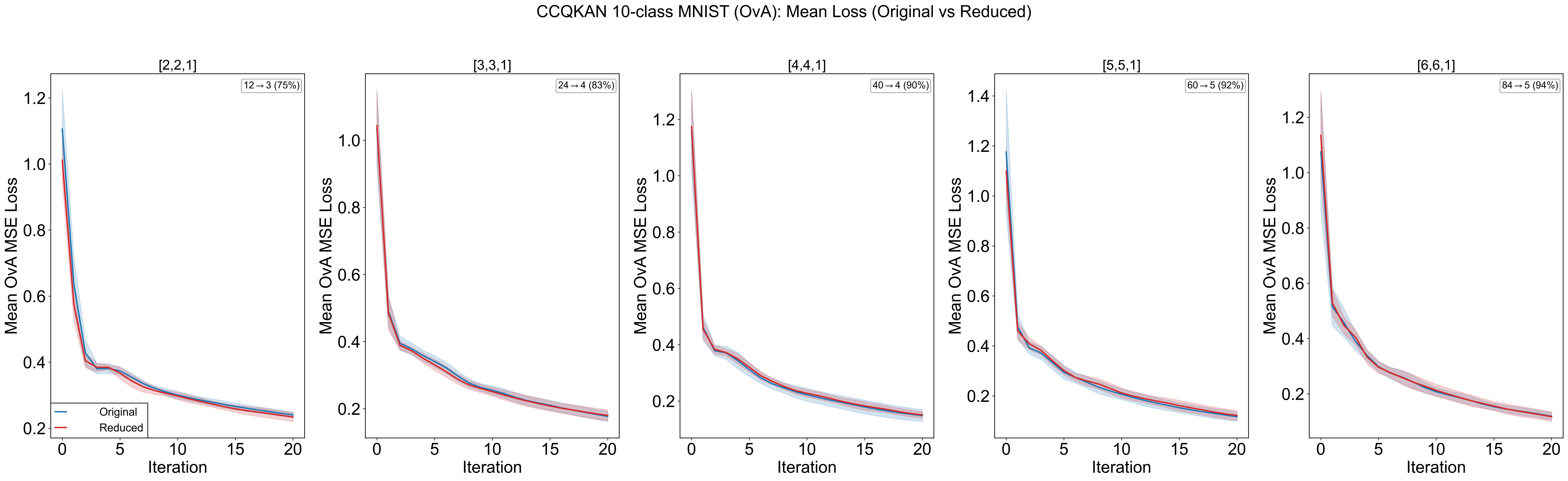}
  \caption{10-class MNIST OvA classification: training loss curves for all five network sizes ($[n,n,1]$, $d=3$, ideal simulation). Solid lines: mean over 10 splits; shaded regions: $\pm 1$ standard deviation. Blue: Original, orange: Red-I. Both models converge at comparable rates.}
  \label{fig:mc_loss}
\end{figure*}

\begin{figure}[!htb]
  \includegraphics[width=\columnwidth]{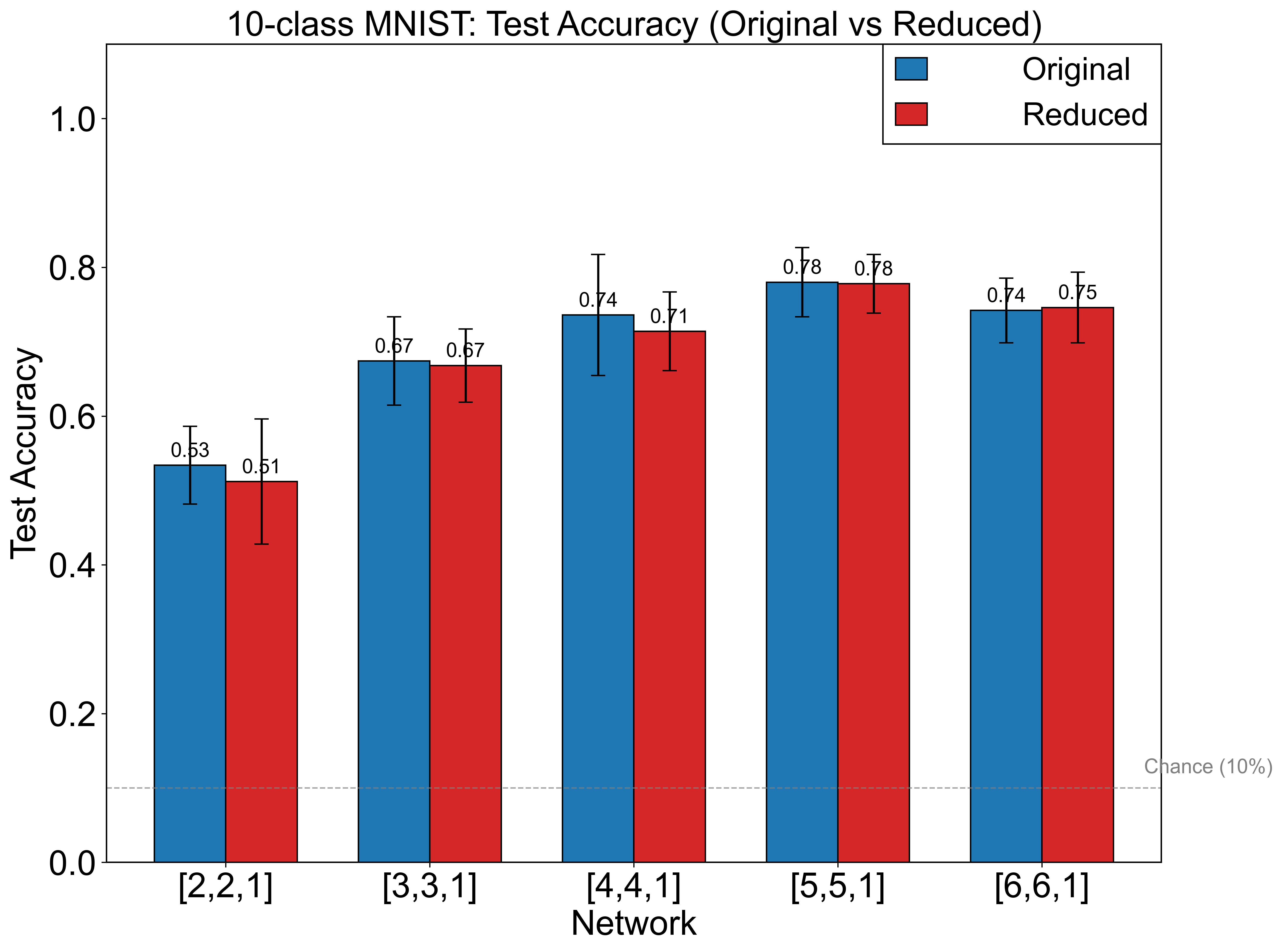}
  \caption{10-class MNIST OvA test accuracy across network sizes. Error bars: $\pm 1$ standard deviation over 10 splits. All accuracies are well above the 10\% chance level (dashed line). No significant difference between Original and Red-I in any configuration.}
  \label{fig:mc_accuracy}
\end{figure}

\begin{figure*}[!htb]
  \includegraphics[width=\textwidth]{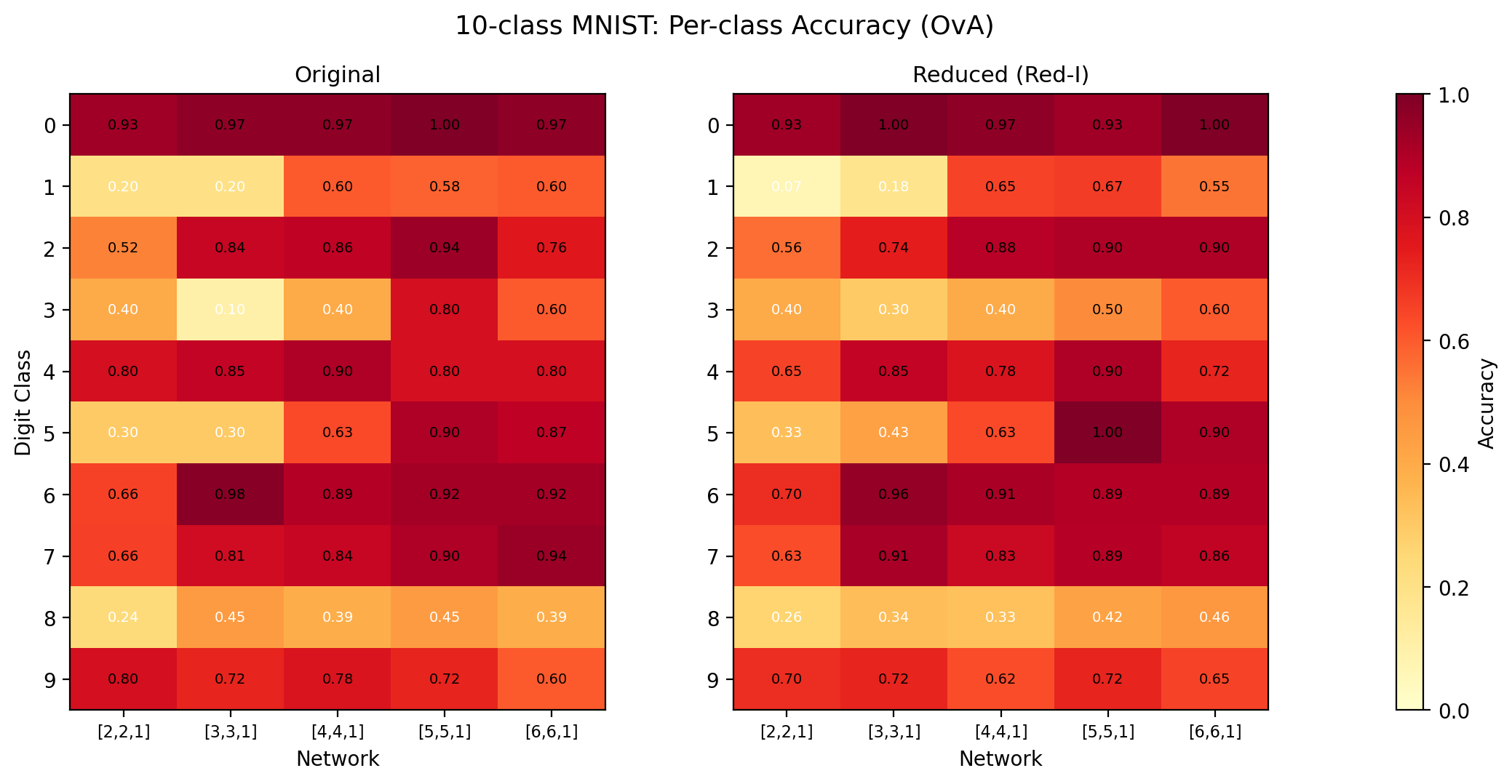}
  \caption{Per-class test accuracy for 10-class MNIST OvA classification (mean over 10 splits). Rows: digit classes 0--9. Columns: Original and Red-I for each network size. Digit~0 is consistently easy; digits~1, 3, and~8 are the most challenging.}
  \label{fig:mc_perclass}
\end{figure*}

Figure~\ref{fig:mc_loss} shows the training loss curves for all five network sizes.
Original and Red-I converge at comparable rates and to similar final losses across all configurations, mirroring the behavior observed in the synthetic experiments under ideal conditions.
Figure~\ref{fig:mc_accuracy} compares overall test accuracy, while Fig.~\ref{fig:mc_perclass} shows per-class accuracy across all configurations.
Per-class analysis reveals that digit~0 is consistently the easiest to classify (93--100\%), while digits~1, 3, and~8 are the most challenging (7--46\% for $n=2$--$3$).
Neither architecture shows a systematic per-class advantage; some digits favor Original, others favor Red-I, consistent with random initialization effects.

\begin{table}[!htb]
\caption{10-class MNIST OvA classification results (mean $\pm$ std over 10 random splits, $d=3$, ideal simulation). Only Original and Red-I are compared. Classical baselines (logistic regression and linear SVM) on the same PCA-reduced features are shown for reference.}
\label{tab:mc_mnist}
\begin{tabular}{lccc}
 & Original & Red-I & LinSVC \\
\hline
\multicolumn{4}{c}{Test accuracy} \\
\hline
$[2,2,1]$ & $0.534 \pm 0.052$ & $0.512 \pm 0.084$ & $0.452 \pm 0.040$ \\
$[3,3,1]$ & $0.674 \pm 0.059$ & $0.668 \pm 0.049$ & $0.560 \pm 0.025$ \\
$[4,4,1]$ & $0.736 \pm 0.081$ & $0.714 \pm 0.053$ & $0.710 \pm 0.039$ \\
$[5,5,1]$ & $0.780 \pm 0.047$ & $0.778 \pm 0.039$ & $0.760 \pm 0.046$ \\
$[6,6,1]$ & $0.742 \pm 0.043$ & $0.746 \pm 0.047$ & $0.774 \pm 0.037$ \\
\hline
\multicolumn{4}{c}{Final mean OvA MSE loss} \\
\hline
$[2,2,1]$ & $0.240 \pm 0.009$ & $0.234 \pm 0.013$ & -- \\
$[3,3,1]$ & $0.177 \pm 0.015$ & $0.180 \pm 0.016$ & -- \\
$[4,4,1]$ & $0.148 \pm 0.021$ & $0.151 \pm 0.012$ & -- \\
$[5,5,1]$ & $0.117 \pm 0.019$ & $0.122 \pm 0.020$ & -- \\
$[6,6,1]$ & $0.119 \pm 0.014$ & $0.118 \pm 0.017$ & -- \\
\end{tabular}
\end{table}

\begin{table}[!htb]
\caption{Wilcoxon signed-rank test for 10-class MNIST (Original vs Red-I, 10 splits). All comparisons are non-significant. With $n=10$ paired samples, the test has $\sim$60\% power at $\alpha=0.05$ for a large effect ($r \geq 0.7$).}
\label{tab:mc_wilcoxon}
\begin{tabular}{lcclcc}
 & \multicolumn{2}{c}{Test accuracy} & & \multicolumn{2}{c}{Final loss} \\
\cmidrule(lr){2-3}\cmidrule(lr){5-6}
Network & $p$ & Sig. & & $p$ & Sig. \\
\hline
$[2,2,1]$ & 0.680 & n.s. & & 0.160 & n.s. \\
$[3,3,1]$ & 0.641 & n.s. & & 0.557 & n.s. \\
$[4,4,1]$ & 0.438 & n.s. & & 0.625 & n.s. \\
$[5,5,1]$ & 0.879 & n.s. & & 0.695 & n.s. \\
$[6,6,1]$ & 0.898 & n.s. & & 0.770 & n.s. \\
\end{tabular}
\end{table}

\FloatBarrier

\FloatBarrier

\FloatBarrier

\section{Discussion}
\label{sec:discussion}

\subsection{Functional equivalence and expressiveness}
\label{sec:expressiveness}

The contribution of this work is empirical rather than foundational: as established in Sec.~\ref{sec:derivation}, the merged encoding computes the same mathematical function as the original [Eqs.~\eqref{eq:sum_identity}--\eqref{eq:Sj_recovery}], an identity that is mathematically elementary.

We note, however, that functional equivalence provides a stronger theoretical basis for trainability than is typical in quantum architecture comparisons.
Since both circuits evaluate the same function $\hat{y}(\bm{\theta})$ over the same parameter space $\bm{\theta}$, the loss landscape $\mathcal{L}(\bm{\theta})$ is \emph{identical} between the two architectures---including all global and local minima, saddle points, and (in principle) the Hessian structure, though we do not compute the Hessian explicitly.
Under ideal (statevector) simulation with numerical gradients [Eq.~\eqref{eq:numerical_gradient}], the gradient $\partial\mathcal{L}/\partial\theta_j$ depends only on function values $\mathcal{L}(\bm{\theta}\pm\epsilon\bm{e}_j)$, which are the same for both circuits.
Therefore, given identical initialization, the two architectures would follow \emph{exactly} the same optimization trajectory in ideal conditions.
The empirical non-significance of Original vs.\ Red-I under ideal conditions is thus expected on theoretical grounds; our experiments serve as a consistency check confirming this reasoning.

The theoretical argument does not extend to shot-noise or device-noise conditions, where the merged and original circuits produce different measurement statistics due to their different qubit counts and circuit depths.
Under finite-shot estimation, the variance of the gradient depends on circuit-specific measurement probabilities, so the two architectures may exhibit different gradient noise even though they compute the same function in expectation.
Characterizing this noise-gradient interaction analytically is an open problem, analogous to the difficulty of proving absence of barren plateaus for general parameterized circuits~\cite{McClean2018}.
Theoretical trainability guarantees remain elusive even for standard variational quantum circuits~\cite{Cerezo2021}; the present work cannot be expected to resolve this broader open question.
Our empirical results under noisy conditions (Sec.~\ref{sec:results})---consistent across 30 comparisons (see Sec.~\ref{sec:stat_testing} for independence caveats)---provide evidence that the merged encoding does not introduce measurable degradation at the achievable statistical power ($n=16$).

\begin{figure*}[!htb]
  \includegraphics[width=\textwidth]{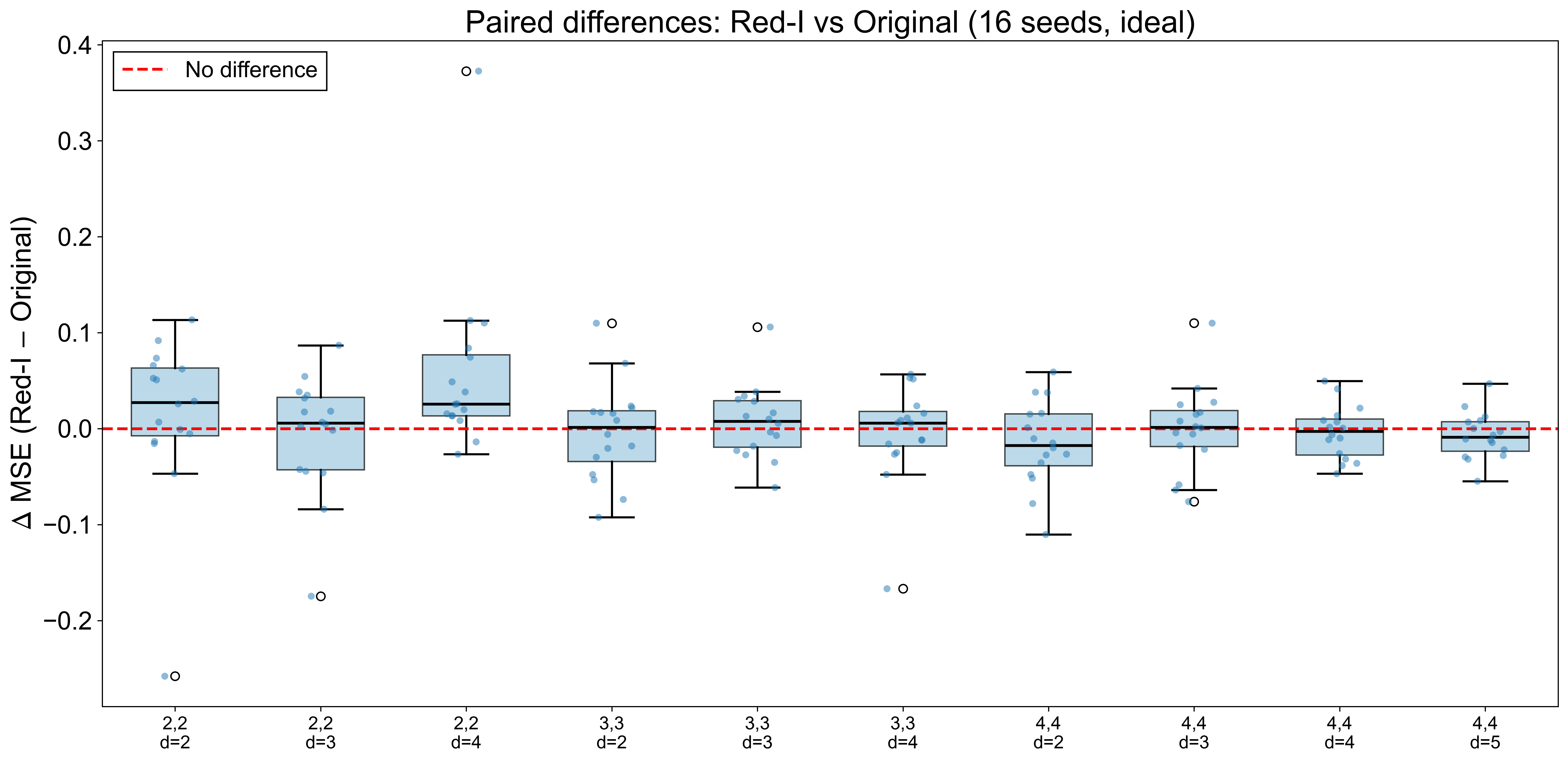}
  \caption{Paired differences in final MSE (Red-I $-$ Original) under ideal conditions (16 seeds). Box plots show the distribution of differences for each of the 10 configurations. The dashed red line marks zero (no difference). Distributions are centered near zero with no systematic bias, supporting the claim that the merged encoding preserves trainability.}
  \label{fig:paired_diff}
\end{figure*}

Figure~\ref{fig:paired_diff} shows the paired differences (Red-I $-$ Original) in final MSE across all 10 configurations under ideal conditions; the distributions are centered near zero with no systematic bias in either direction.

\subsection{Comparison with data re-uploading}
\label{sec:reuploading}

Data re-uploading circuits~\cite{PerezSalinas2020} achieve universal classification by repeatedly encoding classical data into a single-qubit circuit, achieving high expressiveness with minimal qubits.
The merged encoding shares the conceptual principle of packing more classical information into fewer qubits, but differs in mechanism: rather than sequential re-encoding through repeated layers, the merged approach concatenates pre-computed products into a single amplitude state.

To make the comparison quantitative, we trained a parameter-matched single-qubit data re-uploading regressor (layers of $R_y(b_\ell + \bm{w}_\ell \cdot \bm{x})$ rotations with $\langle Z \rangle$ readout, layer count chosen so that the parameter total equals the corresponding CCQKAN's) on the same regression tasks, seeds, and optimization protocols (Table~\ref{tab:reup}).
The merged CCQKAN attains 5--30$\times$ lower final MSE in every configuration and condition (all Wilcoxon $p < 10^{-4}$), while the re-uploading circuit uses a single qubit and a single circuit execution per forward pass.
The two approaches thus occupy complementary resource regimes: re-uploading minimizes width and executions at the cost of accuracy on these targets, whereas the merged CCQKAN buys accuracy with a logarithmic-width register and $n{+}1$ executions.

\begin{table}[!htb]
\caption{Parameter-matched comparison with single-qubit data re-uploading~\cite{PerezSalinas2020} (16 seeds; ideal: 20 Adam steps; shots: SPSA at matched measurement budget, 1000 shots). $|\bm{\theta}|$ is identical for both models in each row.}
\label{tab:reup}
\begin{tabular}{llccc}
Config ($|\bm{\theta}|$) & Condition & Merged MSE & Re-upload MSE & $p$ \\
\hline
$n{=}2$, $d{=}3$ (26)  & ideal & $0.047 \pm 0.024$ & $0.276 \pm 0.057$ & $<10^{-4}$ \\
$n{=}2$, $d{=}3$ (26)  & shots & $0.051 \pm 0.026$ & $0.238 \pm 0.074$ & $<10^{-4}$ \\
$n{=}3$, $d{=}3$ (50)  & ideal & $0.031 \pm 0.012$ & $0.294 \pm 0.085$ & $<10^{-4}$ \\
$n{=}3$, $d{=}3$ (50)  & shots & $0.055 \pm 0.019$ & $0.265 \pm 0.096$ & $<10^{-4}$ \\
$n{=}4$, $d{=}5$ (122) & ideal & $0.013 \pm 0.005$ & $0.386 \pm 0.081$ & $<10^{-4}$ \\
$n{=}4$, $d{=}5$ (122) & shots & $0.030 \pm 0.011$ & $0.269 \pm 0.065$ & $<10^{-4}$ \\
\end{tabular}
\end{table}

Both approaches trade qubits for circuit depth or circuit executions.
Data re-uploading uses $O(1)$ qubits with circuit depth proportional to the number of data re-upload layers; the merged encoding uses $O(\log(nd))$ qubits with depth comparable to a single amplitude encoding.
However, data re-uploading achieves universality through the depth of re-encoding layers, while the merged encoding inherits expressiveness from the original KAN architecture and cannot be directly compared in terms of expressiveness per qubit.
In summary, data re-uploading achieves universality at the cost of circuit depth; the merged encoding achieves circuit-execution reduction at the cost of $O(\log n)$ additional qubits.
A systematic empirical comparison on common benchmarks would clarify when each strategy is preferable, particularly in the context of hardware-specific noise profiles.

\subsection{Parameter transfer vs.\ independent training}

Parameter transfer confers no architectural advantage.
Under ideal conditions, the lower loss of Red-T relative to the 20-step Original (Table~\ref{tab:sig_ideal}) is entirely an optimization-budget effect: the budget-matched control reported with the ideal-condition results shows that an Original trained for 40 steps and a Red-T (20+20) reach identical final losses in every configuration, as the shared loss landscape requires.
Under finite-shot training with the finite-difference estimator of the primary protocol, the additional steps did not improve on the transferred solution; this is consistent with the estimator-breakdown mechanism identified in the finite-shot results rather than with a domain-mismatch effect.
We therefore draw no transfer-learning conclusion from these experiments: parameter transfer is a convenient warm start whose benefit is exactly the extra optimization it inherits.

\subsection{Noise effects}

The progressive equalization of all three models as noise increases (ideal $\to$ shots $\to$ shots + noise; Figs.~\ref{fig:ideal_loss}--\ref{fig:noise_loss} and Table~\ref{tab:sig_summary}) is consistent with the noise-induced flattening of the optimization landscape established for parameterized quantum circuits~\cite{Wang2021noise,Stilck2021}.
When the noise floor exceeds the performance gap between architectures, the choice of circuit becomes secondary to noise mitigation strategies such as zero-noise extrapolation~\cite{Temme2017,Li2017} or increased shot budgets.

Our noise model is simplified: depolarizing noise is applied at the statevector level rather than as gate-level noise with hardware-specific connectivity constraints, crosstalk, readout errors, or $T_1/T_2$ decoherence~\cite{Kjaergaard2020}.
Real NISQ devices exhibit spatially correlated noise whose interaction with the merged encoding may differ from the uniform depolarizing channel studied here.
Therefore, the conclusion that noise equalizes performance across architectures should be understood as applying to the noise model used in this study; validation on actual quantum hardware is necessary before definitive claims about NISQ suitability can be made.
Nevertheless, the statistical equivalence of Red-I and Original under our noise conditions (Table~\ref{tab:sig_summary}) provides evidence that the merged encoding does not introduce additional noise sensitivity.

\subsection{Scaling considerations}

The logarithmic scaling of $Q_{\mathrm{red}}$ versus the quadratic scaling of $Q_{\mathrm{par}}$ implies that the qubit savings relative to the parallel baseline grow with network size.
For illustration (not validated experimentally), extrapolating from Eqs.~\eqref{eq:qubits_orig} and \eqref{eq:qubits_red}, a hypothetical $[10,10,1]$ network with $d=4$ would require $Q_{\mathrm{par}} = 330$ qubits, $Q_{\mathrm{seq}} = 3$ qubits with 110 circuit executions, or $Q_{\mathrm{red}} = 6$ qubits with 11 circuit executions.
However, the sequential nature of the merged evaluation incurs a time overhead proportional to $n$, introducing a qubit--latency trade-off.
On hardware with limited coherence times~\cite{Kjaergaard2020}, this trade-off requires careful characterization.
The merged encoding generalizes straightforwardly to deeper $[n_1, n_2, \ldots, n_L, 1]$ KAN architectures: at each layer, the $n_{\ell}$ input edges for a given output node can be merged into a single circuit execution, independently of other layers.
The circuit execution reduction factor remains $n_{\ell}$ per layer, though the total number of sequential evaluations grows with depth.
However, the reference CCQKAN experiments~\cite{GonzalezOtero2025} use at most two layers ($[1,2,1]$, $[4,1]$, and $[2,1]$ at $d=2$) and our own experiments are confined to $[n,n,1]$; the behavior of deeper architectures---and the interaction of layer-wise merging with depth-dependent gradient issues---remains an open question.

\subsection{Limitations}
\label{sec:limitations}

Several limitations should be noted.

First, all experiments are performed at small scale ($n \leq 4$, $d \leq 5$, $Q_{\mathrm{red}} \leq 5$ qubits) using classical statevector simulation; no quantum advantage is claimed or expected.
The qualitative conclusions should be tested on actual quantum hardware before claims of practical utility can be made.
The smallest non-trivial configuration ($[2,2,1]$, $d=2$, $Q_{\mathrm{red}}=3$) is within reach of current devices (e.g., IBM Eagle/Heron processors with $>100$ qubits); a hardware proof-of-concept comparing sequential and merged executions at matched total shot budgets would provide the most direct validation of our simulation results.
The primary experiments use a 20-step Adam training budget; the loss curves in ideal conditions are still decreasing at iteration 20 (Fig.~\ref{fig:ideal_loss}).
To verify that the conclusions hold beyond early-stage optimization, we conducted supplementary 200-step experiments on two representative configurations ($[2,2,1]\ d\!=\!2$ and $[3,3,1]\ d\!=\!3$) under ideal conditions with 16 seeds.
The final MSE decreased by $\sim$40$\times$ relative to 20 steps ($[2,2,1]\ d\!=\!2$: $0.135 \to 0.003$; $[3,3,1]\ d\!=\!3$: $0.085 \to 0.002$), confirming near-convergence.
The Wilcoxon test remained non-significant in both cases ($p = 0.53$ and $0.98$), with negligible effect sizes ($|r| = 0.19$ and $0.02$).
This confirms that the trainability equivalence persists through convergence, as expected from the identical loss landscape argument (Sec.~\ref{sec:expressiveness}).

Second, the noise model is simplified: depolarization is applied globally to the prepared states, exactly at the level of the SWAP-test statistics, rather than as gate-level noise with hardware-specific connectivity, crosstalk, or readout errors.
Additionally, the sign of the inner product overlap is obtained from the exact statevector rather than from a Hadamard test (see also Sec.~\ref{sec:sign_recovery}), and numerical gradients [Eq.~\eqref{eq:numerical_gradient}] are used instead of the parameter-shift rule~\cite{Mitarai2018,Schuld2019gradients} required on hardware.
On a quantum device, finite differences with $\epsilon = 10^{-5}$ would be overwhelmed by shot noise; the parameter-shift rule evaluates gradients at finite parameter shifts that are robust to measurement noise.
Since the merged encoding alters the circuit structure (amplitude encoding of an $n(d+1)$-dimensional vector versus a $(d+1)$-dimensional vector), the parameter-shift rules would involve different circuit depths and qubit counts, potentially affecting gradient variance.
The SPSA rerun at matched measurement budget provides a first hardware-compatible gradient comparison; an analytic, parameter-shift-style treatment of the gradient variance of the merged circuit remains open.
To probe the intermediate noise regime, we additionally tested $N_{\mathrm{shots}} = 10{,}000$ across all 10 configurations (16 seeds each).
At this higher shot budget, the MSE remains in the 2.2--3.9 range (compared to 2.1--3.9 at 1000 shots); per-configuration tests find 8 of 10 configurations non-significant.
However, because all comparisons address the same directional hypothesis, the appropriate primary analysis is an aggregate test: pooling all 160 seed-level paired differences yields a significant deficit for the merged circuit (Wilcoxon $p = 0.0075$, mean difference $+0.30$ MSE; configuration-level signed Stouffer $Z = +2.74$, $p = 0.006$; configuration-level $t$-test $p = 0.051$), with the merged circuit worse on average in 8 of 10 configurations.
A plausible mechanism is the larger loss-estimate variance of the merged circuit at fixed shot count (the $\|\bm{m}_j\|\sqrt{D}$ amplification of estimator noise grows with $n$).
We therefore explicitly flag this as evidence \emph{against} trainability preservation in the finite-shot regime at this budget, and defer a definitive assessment to experiments with hardware-compatible gradient estimators and matched total-shot budgets.
Still higher budgets ($10^5$--$10^6$) were not tested due to computational cost.
These simplifications mean that the noise-condition results should be interpreted qualitatively.

Third, while the 10-class OvA experiment (Table~\ref{tab:mc_mnist}) provides a more meaningful benchmark than the binary task, both the test set size ($N_{\mathrm{test}} = 50$) and class imbalance (digit~3 has only 1 test sample) limit the reliability of per-class accuracy estimates.
Extension to higher-resolution datasets and larger test sets would strengthen the conclusions.
Additionally, the use of MSE rather than cross-entropy loss for classification follows the reference CCQKAN protocol~\cite{GonzalezOtero2025} but is suboptimal for classification tasks; the choice of loss function may interact differently with the merged encoding and warrants investigation.

Fourth, the total gate count per forward pass is comparable between merged and sequential approaches (Sec.~\ref{sec:resource}), so the practical value of the $n$-fold circuit execution reduction [Eq.~\eqref{eq:circuit_reduction}] depends on hardware-specific costs that we have not benchmarked.
In particular, amplitude state preparation for the merged vector ($n(d+1)$ components) requires $O(nd)$ CNOT gates per execution~\cite{Shende2006}, compared to $O(d)$ for each sequential execution.
If per-circuit overhead (job scheduling, classical control, readout) dominates, the merged approach offers a wall-clock advantage; if gate-level noise or circuit depth is the bottleneck, the advantage may be negated.
We have not measured wall-clock times in simulation because statevector simulation cost scales differently from hardware execution cost; a meaningful wall-clock comparison would require benchmarking on actual quantum hardware or calibrated hardware emulators.

Fifth, in the simulation protocol the element-wise products $\bm{c}_{ij} \odot \bm{T}(x_i)$ that populate the merged state, its norm $\|\bm{m}_j\|$, and the sign of $S_j$ are all computed classically before the circuit is prepared.
The quantum subroutine therefore estimates a quantity whose value is already available classically: the merged encoding relocates the multiplications from the quantum register (where the original CCQKAN performs them implicitly through the inner product of separately encoded $\bm{c}$ and $\bm{T}$) to classical preprocessing.
Any end-to-end computational benefit would have to come from a setting in which $\bm{T}(x_i)$ or the products are prepared quantumly, or from the per-execution overheads discussed above; we make no claim of such a benefit here.

Sixth, the $n$-fold reduction in circuit executions does not translate into a reduction in total measurement cost at fixed estimation precision.
The all-zero outcome probability of the merged circuit scales as $S_j^2/(D\|\bm{m}_j\|^2) \propto 1/D$ for inputs whose summands partially cancel, so each merged execution requires $O(n)$ more shots than a sequential execution for the same relative precision, and the total shot count of both schemes scales as $O(n^2)$ per layer.
The execution reduction is thus a reduction in the number of circuit submissions, not in measurement shots, which is why the finite-shot comparison was performed at matched total measurement budget.

\section{Conclusion}
\label{sec:conclusion}

Merged amplitude encoding reduces the circuit executions of CCQKAN by a factor of $n$ using no more qubits than the sequential SWAP-test baseline, without measurably degrading trainability under ideal conditions---the central empirical finding of this work.
Concretely, the method computes the sum of $n$ edge activations in a single circuit execution using $\lceil\log_2(n(d+1))\rceil$ qubits and is mathematically equivalent to the original sequential evaluation [Eqs.~\eqref{eq:sum_identity}--\eqref{eq:Sj_recovery}].
Our numerical experiments across 10 network configurations, three noise conditions, and 16 random seeds show no statistically significant per-configuration difference between the independently initialized merged circuit and the original in 28 of 30 comparisons (Table~\ref{tab:sig_summary}); the two significant cases are consistent with the nominal Type~I error rate, and we caution that non-significance does not establish equivalence and that the finite-shot conditions have limited discriminative power at this shot budget.
Parameter transfer from a trained Original lowers the final loss relative to a 20-step Original (Table~\ref{tab:sig_ideal}), but this is an optimization-budget effect: at matched budget the two coincide, so transfer confers no architectural advantage.
On binary digit classification, all variants achieve near-perfect accuracy (Table~\ref{tab:mnist}); on the more challenging 10-class OvA task, Original and Red-I achieve comparable test accuracies of 53--78\% with no significant difference detected in any configuration (Table~\ref{tab:mc_wilcoxon}), consistent with the trainability preservation observed in the synthetic experiments.

These findings suggest that merged amplitude encoding can redistribute quantum resources in CCQKAN circuits---reducing circuit executions by a factor of $n$ with no increase in qubit count---without measurably degrading trainability under ideal conditions, and at a small, $n$-dependent cost under finite-shot training.
Key caveats include the small scale ($Q_{\mathrm{red}} \leq 5$, classical simulation only), simplified noise model, and the interpretation of non-significant results as absence of detected difference rather than proof of equivalence (Sec.~\ref{sec:stat_testing}).
Supplementary experiments at higher shot budgets ($N_{\mathrm{shots}} = 10{,}000$), longer training (200 steps to near-convergence), and an SPSA rerun at matched measurement budget together refine this picture: ideal-condition equivalence is exact, while finite-shot training reveals a mild merged-circuit deficit that grows with $n$, quantifying the practical price of the $n$-fold execution reduction.
Validation on actual quantum hardware at larger scale, with hardware-specific noise models and gate-level circuit compilation, remains an important direction for future work, as does extension to more challenging learning tasks where the qubit--execution trade-off may become practically relevant.
The parameter-matched comparison with data re-uploading circuits~\cite{PerezSalinas2020} (Table~\ref{tab:reup}) indicates complementary resource regimes; extending it to other qubit-efficient encodings and to hardware would further clarify when the merged approach is the preferred option.
Despite these limitations, the present results establish a baseline for merged amplitude encoding and provide a concrete, testable prediction for hardware experiments: at matched total shot budget, the merged and sequential circuits should perform equivalently for $n=2$, while for $n \geq 3$ the merged circuit should show a small loss deficit growing with $n$, in exchange for $n$-fold fewer circuit submissions per forward pass.

\bmhead{Acknowledgements}

The author acknowledges the use of cloud computing resources for the numerical simulations reported in this work.

A reference implementation of merged amplitude encoding is provided as the open-source \texttt{qkan} package (Apache-2.0)~\cite{qkanSoftware}, which exposes the original, sequential, and merged CCQKAN circuits behind a single API and reproduces the resource accounting, the merged--sequential equivalence to machine precision, and the ideal-condition experiments.
Reproduction scripts for the finite-shot, noise, and digit-classification experiments will be included in the archived version; until then those results are reproducible from the archived raw data accompanying the manuscript.
The released version will be archived with a permanent DOI upon acceptance.

\bibliography{ccqkan_refs}

\end{document}